\documentstyle[prx,amssymb,prb,aps,pstricks,epsfig,twoside]{revtex}
\input times.sty
\newcounter{puppet}
\newlength{\decade}
\let\rulerson=0
\begin{document}
\title{Local-field study of phase conjugation in metallic quantum
  wells with probe fields of both propagating and evanescent
  character} 
\author{Torsten Andersen\cite{TA:Address} and Ole Keller}
\address{Institute of Physics, Aalborg University,
  Pontoppidanstr{\ae}de 103, DK--9220 Aalborg {\O}st, Denmark} 
\date{Accepted for publication in The Physical Review B} 
\maketitle
\begin{abstract}
  The phase conjugated response from nonmagnetic multi-level metallic
  quantum wells is analyzed and an essentially complete analytical
  solution is presented and discussed. The description is based on a
  semi-classical local-field theory for degenerate four-wave mixing in
  mesoscopic interaction volumes of condensed media developed by the
  present authors [T. Andersen and O. Keller, Phys. Scripta {\bf 58},
  132 (1998)]. The analytical solution is supplemented by a numerical
  analysis of the phase conjugated response from a two-level quantum
  well in the case where one level is below the Fermi level and the
  other level is above. This is the simplest configuration of a
  quantum well phase conjugator in which the light-matter interaction
  can be tuned to resonance. The phase conjugated response is examined
  in the case where all the scattering takes place in one plane, and
  linearly polarized light is used in the mixing. In the numerical
  work we study a two-monolayer thick copper quantum well using the
  infinite barrier model potential. Our results show that the phase
  conjugated response from such a quantum-well system is highly
  dependent on the spatial dispersion of the matter response. The
  resonances showing up in the numerical results are analytically
  identified from the expressions for the linear and nonlinear
  response tensors. In addition to the general discussion of the phase
  conjugated response with varying frequency and parallel component of
  the wavevector, we present the phase conjugated response in the
  special case where the light is in resonance with the interband
  transition.
\end{abstract}

\begin{multicols}{2}
\section{Introduction}
Since the birth of nonlinear optics\cite{Franken:61:1} as a discipline
in physics nonlinear optical processes have been of great interest to
scientists, for instance to help describe surfaces and interfaces of
condensed matter.\cite{Liebsch:97:1,Guyot-Sionnest:87:1,McGilp:96:1}
Studies of surfaces and interfaces of condensed media belong to the
regime of mesoscopic physics, where also quantum wells, -wires and
-dots can be found.\cite{Weisbuch:91:1,Axt:98:1} Among the many
nonlinear optical phenomena that has been studied in the regime of
mesoscopic physics are (i) second harmonic generation from
magnetic\cite{Pan:89:1,Reif:91:1,Rasing:98:1,Vollmer:98:1} as well as
nonmagnetic\cite{Sipe:82:1,Richmond:88:1,Liebsch:89:1,Heinz:91:1,%
  Janz:93:1,Reider:95:1,Liebsch:95:1,Pedersen:95:1,Liu:99:2} systems,
(ii) sum- and difference frequency generation,\cite{Reider:95:1,%
  Bavli:91:1} of which one of the most prominent applications today is
Sisyphus cooling of atoms,\cite{Chu:98:1,Phillips:98:1,%
  Cohen-Tannoudji:98:1} (iii) photon drag,\cite{Keller:93:1,%
  Vasko:96:1,Chen:97:1,Keller:97:2} (iv) DC-electric-field induced
second harmonic generation,\cite{Aktsipetrov:95:1,Aktsipetrov:96:1}
(v) the second-order Kerr effect,\cite{Rasing:98:1,Pustogowa:94:1,%
  Liu:95:1,Rasing:95:1,Rasing:96:1,Dewitz:98:1} (vi) electronic and
vibrational surface Raman scattering,\cite{Nkoma:89:1,%
  Mischenko:95:1,Garcia-Vidal:96:1} (vii) two-photon
photoemission,\cite{Haight:95:1,Fauster:95:1,Georges:95:1,%
  Shalaev:96:1,Tergiman:97:1,Wolf:99:1,Tomas:99:1} (viii) generation
of higher harmonics,\cite{vonderLinde:96:1,Garvila:92:1} (ix) the
second-order Lorenz-Mie scattering,\cite{Dewitz:96:1} and (x)
degenerate four-wave mixing.\cite{Chang:98:1,Andersen:98:2,Liu:99:1,%
  Goldstein:99:1,Deng:99:1}

In the present communication we study phase conjugation by degenerate
four-wave mixing in a quantum-well structure, where both interband and
intraband transitions are allowed. Phase conjugation is a nonlinear
process where the response field is counterpropagating to an incoming
probe field. The usual descriptions (see
Refs.~\onlinecite{Fisher:83:1,Zeldovich:85:1,Pepper:85:1,Sakai:92:1,%
  Gower:94:1} and references therein) of degenerate four-wave mixing
(DFWM) are based on the assumption that the field amplitudes are
slowly varying on the optical wavelength scale [slowly varying
envelope (SVE) approximation], and thus also on the electric dipole
(ED) approximation. We have previously presented the reasons (in
Refs.~\onlinecite{Andersen:97:1} and \onlinecite{Andersen:98:2}) why
these approximations are invalid when considering optical interactions
with matter of mesoscopic size, especially when evanescent components
of the optical field are present. Four-wave mixing in media with
two-dimensional translational invariance has so far been studied by
other authors in the context of phase conjugation of electromagnetic
surface waves,\cite{Fukui:78:1,Ujihara:82:1,Ujihara:82:2} and of a
bulk wave by surface waves.\cite{Zeldovich:80:1,Ujihara:83:1,%
  Stegeman:83:1,Ninzi:84:1,Nunzi:84:2,Mamaev:84:1,Mukhin:85:1,%
  Arutyunyan:87:1,Pilipetskii:87:1} In these investigations
macroscopic electrodynamic approaches were used. In order to go beyond
the SVE and ED approximations a nonlocal microscopic theoretical model
for optical phase conjugation by DFWM has been constructed (see
Ref.~\onlinecite{Andersen:97:1}) for nonmagnetic media. In addition to
avoiding the SVE and ED approximations, other usually made
approximations when considering optical phase conjugation are avoided
in our model, namely (i) the paraxial approximation, (ii) the
assumption of a lossless medium, (iii) the assumption of a weak probe
field, and (iv) the requirement of phase matching between the
interacting fields.

In a two-dimensional translationally invariant system the change in
energy of an electron due to an electric field can either involve a
change of momentum along the translationally invariant plane
(intraband transition), a change of energy eigenstate perpendicular to
the translationally invariant plane (interband transition), or both.
The change of momentum along the translationally invariant plane
occurs as an addition of the momentum parallel to the plane of the
interacting field component (denoted by $\vec{q}_{\|}$) to (or
subtraction from) the momentum of the electron parallel to the surface
(denoted $\vec{\kappa}_{\|}$). It is thus convenient to divide the
photon momentum $\vec{q}$ into its components parallel and
perpendicular to that plane, i.e., $\vec{q}=(\vec{q}_{\|},q_{\perp})$.
Then the vacuum dispersion relation $\vec{q}\cdot\vec{q}=q^2=
\omega^2/c_0^2$ provides us with an extra degree of freedom, since
$q_{\|}=|\vec{q}_{\|}|$ can be larger than $\omega/c_0$. Using the
vacuum dispersion relation we find that $q_{\perp}$ becomes imaginary
in that case. In the following, propagating field components thus
shall refer to the case where $q_{\perp}$ is a real quantity, and
evanescent field components to the case where $q_{\perp}$ is a purely
imaginary quantity. If we want to get a broad understanding of the
phase conjugated response of a probe containing both propagating and
evanescent field components from a quantum-well phase conjugator, two
cases are of fundamental interest, namely (i) the pure intraband case
and (ii) the case where also an interband transition is involved.

The phase conjugated response from a pure intraband quantum well we
have described in Ref.~\onlinecite{Andersen:98:2}. This analysis
revealed that the phase conjugation reflection coefficient is not only
highly nonuniform in the $\vec{q}_{\|}$-spectrum, but also that the
coupling efficiency is several orders of magnitude larger in part of
the evanescent regime than in the propagating regime. Since evanescent
waves are strongly decaying in space we further concluded that if one
wants to see the phase conjugation of evanescent modes, both
excitation and observation should take place close to the surface of
the quantum well. Furthermore was discussed the problems of excitation
of the near-field regime, and the consequences from choosing a
broadbanded (with respect to $\vec{q}_{\|}$) two-dimensional point
source (quantum wire) revealed that parts of the evanescent spectrum
could be excited, and in Ref.~\onlinecite{Andersen:98:3} that phase
conjugation of evanescent fields can lead to a focus of the phase
conjugated field substantially below the so-called diffraction
limit.\cite{Abbe:73:1,Rayleigh:96:1} Since this has also been
experimentally observed,\cite{Bozhevolnyi:94:1} we judge that it is
highly relevant also to give an account of how evanescent fields are
phase conjugated in a system where not only one electronic level is
present.

Since including more than one interband transition will be necessary
for most practical applications, we present in this paper the complete
solution to the theoretical model of Ref.~\onlinecite{Andersen:97:1}
in the case of two-dimensional translational invariance, although a
description based on the self-field approximation according to the
Feibelman theory\cite{Feibelman:75:1,Feibelman:82:1} would be
sufficient in order to determine the dominating response. Giving a
complete solution also allows us to comment on what we would lose
using the self-field approximation. The solution is based on a
discretization in the energy levels of the two-dimensionally
tranlational invariant medium.  Contrary to discretization schemes
performed in real space or Fourier space, our discretization does not
in itself imply an approximation.  Thus, once the complete orthonormal
set of wavefunctions for the phase conjugating medium has been
determined, the phase conjugated response can in principle be
calculated from the solution presented in this paper. How to find the
proper set of wavefunctions for a given material system is another
problem, which for example can be treated using one of several
band-structure methods,\cite{Holzwarth:97:1,Krasovskii:97:1} e.g., the
Korringa-Kohn-Rostoker (KKR),\cite{Korringa:47:1,Kohn:54:1} the
linearized augmented plane-wave (LAPW),\cite{Singh:94:1} or the linear
muffin-tin orbital (LMTO) method.\cite{Andersen:75:1} These methods
are based on an atomic description of the potential in a certain
radius of each atom, adding exchange- and correlation
terms\cite{Kohn:65:1} and different approximations in the regions
between the atomic boundaries. Using such a method one will probably
be able to give more accurate numerical results for specific
materials, but at the cost of the (relatively) analytical simplicity.
Therefore, we will not elaborate further on this point here, but in
stead resort to using a simple description of the matter
wavefunctions. Doing so, we will be able to present a qualitative
discussion based on analytical expressions.

Using a two-level quantum-well phase conjugator, it is also possible
to study resonant four-wave mixing, which until now has been studied
only without spatial dispersion (at the point $(q_{\|},\omega)=
(0,\omega_{21})$ in the $q_{\|}$-$\omega$-plane, $\omega_{21}$ being
the interband transition frequency), as described in, e.g.,
Refs.~\onlinecite{Ducloy:84:1,Pawelek:96:1,Schirmer:97:1,Chalupszak:94:1}.

Thus, in Sec.~\ref{sec:II} we present the theory in the form of a
local-field formalism, we choose a scattering geometry, and the
solution is presented as a discretization in the energy eigenstates.
In Sec.~\ref{sec:III} we prepare for a numerical calculation. We start
by adopting the simple infinite barrier (IB) model to describe the
quantum well.  Furthermore we define the phase conjugation reflection
coefficient, and the section is concluded with a specific choice of a
convenient system to investigate. To give an impression of the
implications of our theoretical model we have presented in
Sec.~\ref{sec:IV} numerical calculations for a two-level quantum-well
phase conjugator. The calculation is supplemented by a discussion of
the results, in particular an identification of the different
resonances appearing when the wavenumber along the surface plane as
well as the frequency varies.  In Sec.~\ref{sec:V} we widen our
discussions, with emphasis on (i) the interband resonance, (ii) the
validity of the self-field approximation, and (iii) the choice of
appropriate relaxation times.  Finally, in Sec.~\ref{sec:VI}, we
conclude.

\section{Theory}\label{sec:II}
As a forerunner for the analysis of the optical phase conjugation from
a two-level quantum well we briefly sketch how a local-field
calculation allows one to determine the so-called degenerate four-wave
mixing response of a mesoscopic metallic film deposited on a
dielectric substrate. To create a phase conjugated field, which in the
plane of the film propagates in a direction opposite to that of the
probe field, two counterpropagating pump fields must be present inside
the phase conjugating medium. Although the theoretical model developed
in Ref.~\onlinecite{Andersen:97:1} allows us to make almost arbitrary
choices of the interacting optical fields, we will in the present work
assume for simplicity that the pump fields (i) propagate parallel to
the plane of the film, and (ii) have constant amplitude across the
film. The scattering geometry is shown in Fig.~\ref{fig:1} together
with the chosen coordinate system. We will further limit our study to
the case where (iii) scattering takes place in the $x$-$z$-plane, and
(iv) the interacting fields are linearly polarized, either in
(p-polarized) or perpendicular to (s-polarized) the scattering plane.
Since it is necessary in a study of nonlinear optical phenomena in
mesoscopic interaction volumes to abandon macroscopic electrodynamics,
the starting point is the microscopic Maxwell-Lorentz equations. The
phase conjugated field from a quantum well exhibiting
free-electron-like dynamics in the plane of the well ($x$-$y$-plane)
can then be described using the single-coordinate ($z$) loop
equation\cite{Keller:96:1}
\begin{intextfigure} 
\setlength{\unitlength}{1mm}
\psset{unit=1mm}
\begin{center}
\begin{pspicture}(0,0)(85,25)
\put(0,0){
 \psline[linewidth=0.25mm]{-}(2,15)(83,15)
 \psframe[linewidth=0.2mm,fillstyle=crosshatch,hatchwidth=0.2mm,hatchsep=1.8mm](2,0)(83,10)
 \psline[linewidth=2.5mm]{->}(0,7)(30,7)
 \psline[linewidth=1.8mm,linecolor=white]{->}(0.35,7)(29,7)
 \put(24,7){\makebox(0,0)[c]{2}}
 \psline[linewidth=2.5mm]{->}(85,7)(55,7)
 \psline[linewidth=1.8mm,linecolor=white]{->}(84.65,7)(56,7)
 \put(61,7){\makebox(0,0)[c]{1}}
 \rput[br]{-27}(15,24){%
  \psline[linewidth=1.5mm]{->}(0,0)(20,0)
  \psline[linewidth=1mm,linecolor=white]{->}(0.25,0)(19.25,0)
  \put(0,-1){\makebox(0,0)[tl]{probe}}
 }
  \psframe[linewidth=0mm,linecolor=white,fillstyle=solid,fillcolor=white](69.25,0.25)(82.75,3.25)
  \put(82.25,0.25){\makebox(-12.5,3)[c]{substrate}}
  \put(82.75,14.5){\makebox(0,0)[tr]{quantum well}}
  \put(82.75,16){\makebox(0,0)[br]{vacuum}}
}
\put(36,0){%
 \pscircle[linewidth=0mm,linecolor=white,fillstyle=solid,fillcolor=white](3,10){1.75}
 \psline[linewidth=1.35mm,linecolor=white]{-}(-0.5,9.875)(15.5,9.875)
 \psline[linewidth=1.35mm,linecolor=white]{-}(3,25)(3,-1)
 \psline[linewidth=1.35mm,linecolor=white]{-}(1.5,15)(4.5,15)
 \pspolygon[linewidth=0mm,linecolor=white,fillstyle=solid,fillcolor=white](12,8.375)(15.5,9.375)(12,9.375)
 \pspolygon[linewidth=0mm,linecolor=white,fillstyle=solid,fillcolor=white](-0.5,0.25)(3.625,0.25)(4.625,3)(-0.5,3)
 \psline[linewidth=0.35mm]{->}(0,9.875)(15,9.875)
 \psline[linewidth=0.35mm]{->}(3,25)(3,0)
 \pscircle[linewidth=0.25mm,linecolor=black,fillstyle=solid,fillcolor=white](3,10){1.0}
 \qdisk(3,10){0.25}
 \put(15,11){\makebox(0,0)[rb]{$x$}}
 \put(2,11){\makebox(0,0)[rb]{$y$}}
 \put(1.75,0.75){\makebox(0,0)[rb]{$z$}}
 \psline[linewidth=0.35mm]{-}(2,15)(4,15)
 \put(3.5,15.5){\makebox(0,0)[bl]{$z=-d$}}
}
\end{pspicture}
\end{center}
\fcaption{The system we consider here consists of a three layer
  structure, namely (i) vacuum, extending from $-\infty$ to $-d$, (ii)
  quantum well, extending from $-d$ to $0$, and (iii) substrate
  (crosshatched), extending from $0$ to $+\infty$. The three incoming
  electromagnetic fields consists of two pump fields (labeled ``1''
  and ``2'') and a probe field. Also shown is the Cartesian coordinate
  system used in our calculations.\label{fig:1}}
\end{intextfigure}
\begin{eqnarray}
 \vec{E}_{\rm{PC}}(z;\vec{q}_{\|},\omega)&=&
 \vec{E}_{\rm{PC}}^{\rm{B}}(z;\vec{q}_{\|},\omega)
 -{\rm{i}}\mu_{0}\omega\int\int
 \tensor{G}(z,z'';\vec{q}_{\|},\omega)
\nonumber\\ &&
 \cdot\tensor{\sigma}(z'',z';\vec{q}_{\|},\omega)\cdot
 \vec{E}_{\rm{PC}}(z';\vec{q}_{\|},\omega)dz''dz',
\label{eq:1}
\end{eqnarray}
where $\omega$ is the common angular frequency of the participating
fields, and $\vec{q}_{\|}$ is the component of the probe field in the
film plane. It is the so-called background field,
$\vec{E}_{\rm{PC}}^{\rm{B}}(z;\vec{q}_{\|},\omega)$, which makes the
loop problem different for the various nonlinear (and linear)
problems. It is here given by
\begin{equation}
 \vec{E}_{\rm{PC}}^{\rm{B}}(z;\vec{q}_{\|},\omega)=-{\rm{i}}\mu_{0}\omega\int
 \tensor{G}(z,z';\vec{q}_{\|},\omega)\cdot
 \vec{J}_{-\omega}^{\,(3)}(z';\vec{q}_{\|},\omega)dz',
\label{eq:2}
\end{equation}
where $\vec{J}_{-\omega}^{\,(3)}(z';\vec{q}_{\|},\omega)$ is the
current density driving the nonlinear process. The pseudo-vacuum
propagator $\tensor{G}(z,z'';\vec{q}_{\|},\omega)$ is given by
\begin{eqnarray}
\lefteqn{
 \tensor{G}(z,z';\vec{q}_{\|},\omega)=
 {e^{{\rm{i}}q_{\perp}|z-z'|}\over2{\rm{i}}q_{\perp}}\left[
 \vec{e}_{y}\otimes\vec{e}_{y}+\Theta(z-z')\vec{e}_{i}\otimes\vec{e}_{i}
\right.}\nonumber\\ &&\left.
 +\Theta(z'-z)\vec{e}_{r}\otimes\vec{e}_{r}\right]
 +{e^{-{\rm{i}}q_{\perp}(z+z')}\over2{\rm{i}}q_{\perp}}\left[
 r^s\vec{e}_{y}\otimes\vec{e}_{y}+r^p\vec{e}_{r}\otimes\vec{e}_{i}\right]
\nonumber\\ &&
 +{1\over{}q^2}\delta(z-z')\vec{e}_{z}\otimes\vec{e}_{z},
\label{eq:5}
\end{eqnarray}
where the first term describes the direct propagation of the
electromagnetic field from a source plane at $z'$ to the observation
plane at $z$, the second term accounts for the reflection at the
quantum-well/substrate interface, and the third term characterizes the
field generated at the observation plane by the current density
prevailing in the same plane (thus named the self-field term).
Above, $\vec{e}_{i}=q^{-1}(q_{\perp},0,-q_{\|})$, and
$\vec{e}_{r}=q^{-1}(-q_{\perp},0,-q_{\|})$, taking
$\vec{q}_{\|}=q_{\|}\vec{e}_{x}$. The quantities $r^s$ and $r^p$ are
the amplitude reflection coefficients at the vacuum/substrate
interface in the absence of the quantum well for s- and
p-polarized fields, respectively. Both of these are in general
functions of ${q}_{\|}$. Moreover, the vectors $\vec{e}_{x}$,
$\vec{e}_{y}$, and $\vec{e}_{z}$ are unit vectors along the principal
axes in the Cartesian $x$-$y$-$z$-coordinate system, $\Theta(\cdots)$
is the Heaviside unit step function, and $\delta(\cdots)$ is the Dirac
delta function.

The $ij$'th tensor element of the
linear response tensor $\tensor{\sigma}(z'',z';\vec{q}_{\|},\omega)$,
appearing in Eq.~(\ref{eq:1}), is given
by\cite{Keller:96:1,Andersen:98:1} \widetext
\begin{equation}
 {\sigma}_{ij}(z,z';\vec{q}_{\|},\omega)=
 {2i\over\hbar\omega}{1\over(2\pi)^2}\sum_{nm}\int{\omega\over
  \tilde{\omega}_{nm}(\vec{\kappa}_{\|}+\vec{q}_{\|},\vec{\kappa}_{\|})}
 {f_{n}(\vec{\kappa}_{\|}+\vec{q}_{\|})-f_{m}(\vec{\kappa}_{\|})\over
  \tilde{\omega}_{nm}(\vec{\kappa}_{\|}+\vec{q}_{\|},\vec{\kappa}_{\|})-\omega}
 j_{i,nm}(z;2\vec{\kappa}_{\|}+\vec{q}_{\|})
 j_{j,mn}(z';2\vec{\kappa}_{\|}+\vec{q}_{\|})
 d^2\kappa_{\|},\!\!\!\!
\label{eq:7}
\end{equation}
provided the set of wavefunctions is complete. In Eq.~(\ref{eq:7})
we have introduced the transition current density in the mixed Fourier
space, namely
\begin{eqnarray}
 \vec{\jmath}_{nm}(z;\vec{Q}_{\|})&=&
 -{e\hbar\over2im_{e}}
 \biggl[i\vec{Q}_{\|}\psi_{m}^{*}(z)\psi_{n}(z)
 +\vec{e}_{z}\left(
 \psi_{m}^{*}(z){d\psi_{n}(z)\over{}dz}
 -\psi_{n}(z){d\psi_{m}^{*}(z)\over{}dz}\right)\biggr].
\label{eq:8}
\end{eqnarray}
\narrowtext\noindent
In relation to Eq.~(\ref{eq:7}), $\vec{Q}_{\|}$ is equal to
$2\vec{\kappa}_{\|}+\vec{q}_{\|}$, where $\vec{\kappa}_{\|}$ is the
wavevector of the given electron in the plane of the well. The
transition current density also occurs in the nonlinear response
tensor (see Appendix~\ref{app:A}) and in this context various
combinations of $\vec{q}_{\|}$, $\vec{k}_{\|}$, and
$\vec{\kappa}_{\|}$ appear in $\vec{Q}_{\|}$. The quantities
$\psi_{a}$, $a\in\{n,m\}$, are the one-dimensional electronic energy
eigenstates of the quantum well belonging to the $z$-direction, and
they satisfy the field-unperturbed Schr{\"o}dinger equation
${\cal{H}}_{0}\psi_a=\varepsilon_a\psi_a$. The quantity
$f_a(\vec{\kappa}_{\|})$ denote the Fermi-Dirac distribution for the
eigenstate
$\Psi_a(\vec{r}\,)=\psi_a(z)\exp(i\vec{\kappa}_{\|}\cdot\vec{r})/(2\pi)$,
where also the solution to the Schr{\"o}dinger equation along the
quantum well is taken into account. It is given by
$f_a(\vec{\kappa}_{\|})=[1+\exp\{(\varepsilon_a
+\hbar^2\kappa_{\|}^2/(2m_e)-\mu)/(k_{\rm{B}}T)\}]^{-1}$, where
$k_{\rm{B}}$ is the Boltzmann constant, $\mu$ is the chemical
potential of the electron system, and $T$ the absolute temperature.
For the various Cartesian components of the transition current
density, we use the notation $j_{i,nm}(z;\vec{\kappa}_{\|})$,
$i\in\{x,y,z\}$. The complex cyclic transition frequency is defined by
\begin{eqnarray}
\lefteqn{
 \tilde{\omega}_{nm}(\vec{Q}_{\|,a},\vec{Q}_{\|,b})=
}\nonumber\\ &\quad&
 {1\over\hbar}\left[\varepsilon_{n}-\varepsilon_{m}
 +{\hbar^2\over2m_{e}}\left(|\vec{Q}_{\|,a}|^2
 -|\vec{Q}_{\|,b}|^2\right)\right]
 -i\tau_{nm}^{-1},
\label{eq:9}
\end{eqnarray}
where $\varepsilon_n$ and $\varepsilon_m$ are the eigenenergies of the
quantum well states belonging to the $z$-direction, and
$\vec{Q}_{\|,a}$ and $\vec{Q}_{\|,b}$ can be any of the relevant
combinations of $\vec{q}_{\|}$, $\vec{k}_{\|}$, and
$\vec{\kappa}_{\|}$. The quantity $\tau_{nm}$ is the relaxation time.

The nonlinear current density,
$\vec{J}_{-\omega}^{\,(3)}(z';\vec{q}_{\|},\omega)$, is related to the
pump and probe fields by a constitutive relation of the form
\begin{eqnarray}
\lefteqn{
 \vec{J}_{-\omega}^{(3)}(z;\vec{q}_{\|},\omega)={1\over(2\pi)^4}\int
 \tensor{\Xi}(z,z';\vec{q}_{\|},\vec{k}_{\|},\omega)
}\nonumber\\ &\quad&
 \vdots
 \vec{E}(-\vec{k}_{\|},\omega)\vec{E}(\vec{k}_{\|},\omega)
 \vec{E}^{*}(z';-\vec{q}_{\|},\omega)dz'+\mbox{i.t.},
\label{eq:6}
\end{eqnarray}
where 
\begin{equation}
 \tensor{\Xi}(z,z';\vec{q}_{\|},\vec{k}_{\|},\omega)=
 \int\int\tensor{\Xi}(z,z',z'',z''';\vec{q}_{\|},\vec{k}_{\|},\omega)
 dz'''dz''
\label{eq:36}
\end{equation}
is the relevant nonlinear response tensor when the pump fields are
essentially constant (slowly varying) across the quantum well, i.e.,
$\vec{E}(z''';-\vec{k}_{\|},\omega)=\vec{E}(-\vec{k}_{\|},\omega)$ and
$\vec{E}(z'';\vec{k}_{\|},\omega)=\vec{E}(\vec{k}_{\|},\omega)$ in
Eq.~(\ref{eq:6}). Within the framework of a single-electron
random-phase-approximation approach an explicit expression for
$\tensor{\Xi}(z,z',z'',z''';\vec{q}_{\|},\vec{k}_{\|},\omega)$ has
been established in Ref.~\onlinecite{Andersen:97:1}. The term ``i.t.''
denotes the so-called ``interchanged term'', which takes into account
the symmetry of the pump fields. It is obtained from the first term by
interchanging the two pump fields (the pump field wavevector
$\vec{k}_{\|}$ is replaced by $-\vec{k}_{\|}$). The explicit
expression for the simplified nonlinear conductivity tensor,
$\tensor{\Xi}(z,z';\vec{q}_{\|},\vec{k}_{\|},\omega)$, can be found in
Appendix~\ref{app:A}, while the more general conductivity tensor
$\tensor{\Xi}(z,z',z'',z''';\vec{q}_{\|},\vec{k}_{\|},\omega)$ has
been given in Ref.~\onlinecite{Andersen:97:1}. We have, however, in
Appendix~\ref{app:A} only listed one of the seven parts, namely part
G, of the nonlinear conductivity tensor that appears in
Ref.~\onlinecite{Andersen:97:1}, since when interband transitions are
strong, it is dominating the response by several orders of magnitude
compared to the other six (A--F).

As a consequence of the above-mentioned choice (but independent of the
direction in which the pump fields propagate) the number of terms in the
nonvanishing elements of the nonlinear response tensor is further
reduced, since the orthonormality of the $z$-dependent parts of the
wavefunction gives
\begin{equation}
 \int\psi_n^*(z)\psi_m(z)dz=\delta_{nm},
\label{eq:37}
\end{equation}
where $\delta_{nm}$ is the Kronecker delta. Also, by integration of
the microscopic transition current density given by Eq.~(\ref{eq:8})
over $z$ one finds
\begin{equation}
 \int{}\vec{\jmath}_{nm}(z;\vec{Q}_{\|})dz
 =-{e\hbar\over{}2im_e}\left[i\vec{Q}_{\|}\delta_{nm}
 +{p}_{z,nm}\vec{e}_{z}\right],
\label{eq:38}
\end{equation}
where
\begin{equation}
 {p}_{z,nm}=\int\left(\psi_{m}^{*}(z){d\psi_{n}(z)\over{}dz}
 -\psi_{n}(z){d\psi_{m}^{*}(z)\over{}dz}\right)dz
\label{eq:39}
\end{equation}
is proportional to the $z$-component of the electric dipole moment
related to the $nm$-transition.\cite{Keller:90:2}

The conductivity tensor
$\tensor{\Xi}(z,z',z'',z''';\vec{q}_{\|},\vec{k}_{\|},\omega)$ has in
general 81 nonzero tensor elements (3$\times$3$\times$3$\times$3) and
consists of seven different parts (A--G) after the seven different
physical processes contributing to the response (see
Ref.~\onlinecite{Andersen:97:1} for details). When scattering takes
place in the $x$-$z$-plane with linearly polarized light the general
treatment can be split into eight separate parts related to the
possible combinations of polarization of the three different incident
fields. In this scattering geometry $\vec{q}_{\|}$ and $\vec{k}_{\|}$
lie along the $x$-axis, giving a mirror plane at $y=0$.  Consequently,
only tensor elements in the nonlinar response tensor with a Cartesian
index even numbered in $y$ contributes, and the $81$ tensor elements
generally appearing are reduced to $41$. The separation of the tensor
elements into the eight sets of elements contributing in these
configurations follows in a straight forward manner from the
definition of the sum-product operator ``$\vdots$'' between the
nonlinear current density and the interacting electric fields, i.e.,
$[\tensor{\Xi}\vdots\vec{E}\vec{E}\vec{E}^{*}]_{i}=
\sum_{jkh}{\Xi}_{ijkh}{E}_{h}{E}_{k}{E}^{*}_{j}$. The added
restriction of letting the pump fields travel along the $x$-axis then
reduces the number of contributing matrix elements from $41$ to $18$,
since when traveling along the $x$-axis, the pump fields are polarized
in either the $y$-direction or the $z$-direction. The resulting sets
of tensor elements we have presented in Table~\ref{tab:I}.

To solve Eq.~(\ref{eq:1}), we can establish a so-called coupled
antenna loop. First, we notice that each matrix element of the linear
conductivity tensor [Eq.~(\ref{eq:7})] with the insertion of
Eq.~(\ref{eq:8}) can be written as a product of a $z$-independent term
and two terms depending on $z$ and $z'$, respectively. Element $ij$
then takes the form
\begin{equation}
 \sigma_{ij}(z,z';\vec{q}_{\|},\omega)=
 \sum_{nm}{\cal{Q}}_{nm}^{ij}(\vec{q}_{\|},\omega)
 {j}_{i,nm}(z){j}_{j,mn}(z'),
\label{eq:10}
\end{equation}

\begin{intexttable}
\tcaption{Contributing tensor elements of the nonlinear conductivity
  tensor when the pump fields are propagating in the $x$-direction and
  all fields are polarized in (p) or perpendicular to (s) the
  $x$-$z$-plane. The left column shows the polarization combination of
  the incoming fields (pump 1, pump 2, probe), the center column shows
  the polarization of the phase conjugated field, and the right column
  shows the tensor elements contributing to the nonlinear
  interaction.\label{tab:I}}
\noindent
\begin{tabular*}{8.6cm}{cc|@{\extracolsep{\fill}}ccc}
\tableline\tableline
input pol. & output pol. & nonlinear tensor elements \\ \tableline
sss & s & $\Xi_{yyyy}$\\
pps & s & $\Xi_{yyzz}$\\
ssp & p & $\Xi_{xxyy}$, $\Xi_{xzyy}$, $\Xi_{zxyy}$, $\Xi_{zzyy}$\\
ppp & p & $\Xi_{xxzz}$, $\Xi_{xzzz}$, $\Xi_{zxzz}$, $\Xi_{zzzz}$\\
spp, psp & s & $\Xi_{yxyz}$, $\Xi_{yxzy}$, $\Xi_{yzyz}$, $\Xi_{yzzy}$\\
sps, pss & p & $\Xi_{xyyz}$, $\Xi_{xyzy}$, $\Xi_{zyyz}$, $\Xi_{zyzy}$\\
\tableline\tableline
\end{tabular*}
\end{intexttable}

\noindent
where $\vec{\jmath}_{nm}(z)\equiv\vec{\jmath}_{nm}(z;\vec{e}_{x}+\vec{e}_{y})$.
The various ${\cal{Q}}$-quantities can readily be identified from
Eq.~(\ref{eq:7}), and the integrals can be solved using the method
described in Appendix~\ref{app:B}. Inserting Eq.~(\ref{eq:10}) into
Eq.~(\ref{eq:1}), we get
\begin{equation}
 \vec{E}_{\rm{PC}}(z)=
 \vec{E}_{\rm{PC}}^{\rm{B}}(z)
 +\sum_{nm} \tensor{F}_{nm}(z)
 \cdot \vec{\Gamma}_{mn},
\label{eq:18}
\end{equation}
omitting the reference to $\vec{q}_{\|}$ and $\omega$ for brevity. In
Eq.~(\ref{eq:18}) we have introduced the $3\times3$ tensor
$\tensor{F}_{nm}(z)$ with the nonzero elements
\begin{eqnarray}
 F_{nm}^{xx}(z)&=&
 -{\rm{i}}\mu_0\omega\sum_{i\in\{x,z\}}
 {\cal{Q}}_{nm}^{xi}\int G_{xi}(z,z''){j}_{i,nm}(z'')dz''
\nonumber\\ &=&
 {q_{\perp}\over{}q_{\|}}F_{nm}^{zx}(z)
,\label{eq:19}\\
 F_{nm}^{xz}(z)&=&
 -{\rm{i}}\mu_0\omega\sum_{i\in\{x,z\}}
 {\cal{Q}}_{nm}^{iz}\int G_{xi}(z,z''){j}_{i,nm}(z'')dz''
\nonumber\\ &=&
 {q_{\perp}\over{}q_{\|}}F_{nm}^{zz}(z)
,\label{eq:20}\\
 F_{nm}^{yy}(z)&=&
 -{\rm{i}}\mu_0\omega{\cal{Q}}_{nm}^{yy}\int G_{yy}(z,z''){j}_{y,nm}(z'')dz'',
\label{eq:21}
\end{eqnarray}
and the elements of the vector $\vec{\Gamma}_{mn}$ are written
\begin{equation}
 {\Gamma}_{i,mn}=\int{j}_{i,mn}(z')E_{{\rm{PC}},i}(z')dz'
,\qquad i\in\{x,y,z\}. \label{eq:24}
\end{equation}
To determine the phase conjugated field the quantity
$\vec{\Gamma}_{mn}$ must be calculated. This is done by multiplication
of each element $E_{{\rm{PC}},i}(z')$, $i\in\{x,y,z\}$ of the phase
conjugated field in Eq.~(\ref{eq:1}) by the relevant ${j}_{i,mn}(z)$
followed by an integration over the $z$-coordinate. Hence, when the
phase conjugated light is s-polarized, Eq.~(\ref{eq:18}) is
transformed into the following set of linear algebraic equations:
\begin{equation}
 \Gamma_{y,mn}-\sum_{vl}K_{yy,mn}^{vl}\Gamma_{y,vl}
 =\Omega_{y,mn},
\label{eq:32}
\end{equation}
i.e., $n^2$ equations with just as many unknowns. In the case of
p-polarized light, we obtain
\begin{eqnarray}
 \Gamma_{x,mn}-\sum_{vl}\left(K_{xx,mn}^{vl}\Gamma_{x,vl}
 +K_{xz,mn}^{vl}\Gamma_{z,vl}\right)&=&\Omega_{x,mn}
,\label{eq:33}\\
 \Gamma_{z,mn}-\sum_{vl}\left(K_{zx,mn}^{vl}\Gamma_{x,vl}
 +K_{zz,mn}^{vl}\Gamma_{z,vl}\right)&=&\Omega_{z,mn}
,\label{eq:34}
\end{eqnarray}
which are $2n^2$ equations with just as many unknowns. In
Eqs.~(\ref{eq:32})--(\ref{eq:34}) above, the elements of the vectorial
quantity $\vec{\Omega}_{mn}$ are given by
\begin{equation}
 {\Omega}_{i,mn}=
 \int{j}_{i,mn}(z)E_{{\rm{PC}},i}^{\rm{B}}(z)dz
,\qquad i\in\{x,y,z\},\label{eq:26}
\end{equation}
and the $3\times3$ tensorial quantity
$\tensor{K}_{mn}^{vl}(\vec{q}_{\|},\omega)$ has the five nonzero
elements
\begin{equation}
 K_{ij,mn}^{vl}=\int{j}_{i,mn}(z)F_{lv}^{ij}(z)dz
,\label{eq:27}
\end{equation}
where the indices ``$i$ and ``$j$'' can take the values of
$ij\in\{xx,xz,yy,zx,zz\}$. By means of the procedure sketched above,
we have been able to transform the integral-equation problem for the
phase conjugated field, $\vec{E}_{\rm{PC}}(z)$, [Eq.~(\ref{eq:1})] to
a matrix problem for the $\vec{\Gamma}_{mn}$-vectors. This
discretization in the energy levels is exact, and once the linear
algebraic set of equations for the $\vec{\Gamma}_{mn}$-vectors,
truncated so as to keep only the subspace of relevant energy levels,
has been solved (numerically) the phase conjugated field can be
obtained from Eq.~(\ref{eq:18}). Integral equations of the type given
in Eq.~(\ref{eq:1}) is often solved (numerically) by discretization in
the real space coordinate. By such a procedure one has to worry about
how small discretization lengths one may dare to take from a physical
point of view. The discretization in energy levels used here does not
suffer from this uncertainty.

\section{Numerical framework}\label{sec:III}
Our description of the phase conjugated field has until now been
independent of the actual wavefunctions in the quantum well, and thus
also independent of the form the potential takes across the active
medium. However, if we want to perform a numerical calculation of the
phase conjugated field we have to choose a definite potential across the
quantum well, giving us a set of wavefunctions to work with. Below we
use the infinite barrier (IB) model potential for the numerical
study, since this model is sufficient for a qualitative study.

As shown in Fig.~\ref{fig:2}, in this model the one-dimensional
potential $V(z)$ is zero inside the quantum well (in the interval
$-d\leq{}z\leq0$) and infinite everywhere else. The stationary state
wavefunctions inside the quantum well are given by
$\psi_{n}(z)=\sqrt{2/d}\sin(n\pi{}z/d)$ and outside the quantum well,
$\psi_n(z)=0$. The associated eigenenergies are
$\varepsilon_n=(n\pi\hbar)^2/(2m_ed^2)$. Within the IB model,
Eq.~(\ref{eq:39}) gives
\begin{equation}
 {p}_{z,nm}={4nm[1-(-1)^{n+m}]\over(n^2-m^2)d}
\label{eq:40}
\end{equation}
for $n\neq{}m$, and $p_{z,nm}=0$ for $n=m$. For a metallic quantum
well one may even at room temperature approximate the Fermi-Dirac
distribution functions by their value at zero temperature, i.e.,
\begin{intextfigure}
\setlength{\unitlength}{1mm}
\psset{unit=1mm}
\begin{center}
\begin{pspicture}(0,0)(80,60)
\psline[linewidth=0.35mm]{<->}(15,50)(15,10)(45,10)(45,50)
\multiput(15,51)(30,0){2}{\makebox(0,0)[b]{$\infty$}}
\psline[linewidth=0.5mm]{-}(14,20)(46,20)
\psline[linewidth=0.5mm,linestyle=dashed,dash=3mm 2mm]{-}(10,30)(50,30)
\psline[linewidth=0.5mm]{-}(14,40)(46,40)
\put(47,20){\makebox(0,0)[l]{$|1\rangle$}}
\put(47,40){\makebox(0,0)[l]{$|2\rangle$}}
\psline[linewidth=0.35mm]{->}(5,10)(5,55)
\psline[linewidth=0.25mm]{-}(4,10)(6,10)
\psline[linewidth=0.5mm]{-}(4,20)(6,20)
\psline[linewidth=0.5mm]{-}(4,30)(6,30)
\psline[linewidth=0.5mm]{-}(4,40)(6,40)
\put(0,10){\makebox(0,0)[l]{$0$}}
\put(0,20){\makebox(0,0)[l]{$\varepsilon_1$}}
\put(0,30){\makebox(0,0)[l]{${\cal{E}}_{\rm{F}}$}}
\put(0,40){\makebox(0,0)[l]{$\varepsilon_2$}}
\put(5,56){\makebox(0,0)[b]{$\varepsilon,V(z)$}}
\psline[linewidth=0.35mm]{->}(0,5)(55,5)
\psline[linewidth=0.25mm]{-}(15,4)(15,6)
\psline[linewidth=0.25mm]{-}(45,4)(45,6)
\put(45,0){\makebox(0,0)[b]{$0$}}
\put(15,0){\makebox(0,0)[b]{$-d$}}
\put(56,5){\makebox(0,0)[l]{$z$}}
\psecurve[linewidth=0.25mm,linestyle=dotted,dotsep=0.5mm]{-}(11.25,36.46)(15,40)(18.75,43.54)(22.5,45)(26.25,43.54)(30,40)(33.75,36.46)(37.5,35)(41.25,36.46)(45,40)(48.75,43.54)
\psecurve[linewidth=0.25mm,linestyle=dotted,dotsep=0.5mm]{-}(7.5,23.54)(15,20)(22.5,16.46)(30,15)(37.5,16.46)(45,20)(52.5,23.54)
\psline[linewidth=0.5mm]{->}(65,20)(65,40)
\psline[linewidth=0.5mm]{->}(75,40)(75,20)
\psline[linewidth=0.25mm]{-}(64,20)(76,20)
\psline[linewidth=0.25mm]{-}(64,40)(76,40)
\pscurve[linewidth=0.5mm]{->}(66,41)(70,43)(74,41)
\pscurve[linewidth=0.5mm]{<-}(66,19)(70,17)(74,19)
\rput[b]{90}(64,30){$\omega_{12},\tau_{12}$}
\rput[b]{270}(76,30){$\omega_{21},\tau_{21}$}
\put(70,44){\makebox(0,0)[b]{$\omega_{22},\tau_{22}$}}
\put(70,16){\makebox(0,0)[t]{$\omega_{11},\tau_{11}$}}
\end{pspicture}
\end{center}
\fcaption{Infinite barrier (IB) model potential (thick solid line) for
  a quantum well with boundaries at $z=-d$ and $z=0$. In the present
  case, only one energy level below the Fermi energy (here called
  $|1\rangle$, with energy $\varepsilon_1$) and one energy level above
  the Fermi energy (called $|2\rangle$, with energy $\varepsilon_2$)
  contributes to the solution. The remaining infinite set of energies
  appearing in the IB model we assume are so far away from $|1\rangle$
  and $|2\rangle$ that they do not contribute to the solution. The
  dotted curves indicate the shape of the wave function for each of
  the two energies. To the right is shown the possible transitions,
  identified with their respective transition frequency and relaxation
  time.\label{fig:2}}
\end{intextfigure}
\begin{equation}
 \lim_{T\rightarrow0}f_n(\vec{\kappa}_{\|})=
 \Theta\left\{{\cal{E}}_{F}-{\hbar^2\over2m_e}
 \left[\left(n\pi\over{}d\right)^2+\kappa_{\|}^2\right]\right\},
\label{eq:41}
\end{equation} 
where ${\cal{E}}_{F}$ is the Fermi energy of the system. In the
low-temperature limit it is possible to find analytical solutions to
the integrals over $\vec{\kappa}_{\|}$ appearing in Eq.~(\ref{eq:A5}).
The explicit calculations are tedious but trivial to carry out, and
since the final expressions are rather long we do not present them
here. For the interested reader some steps in the calculations are
reproduced in Appendix~\ref{app:B}.

The Fermi energy is calculated from the global charge neutrality
condition,\cite{Keller:96:1} and for a quantum well described by the
IB model, it becomes\cite{Andersen:98:1}
\begin{equation}
 {\cal{E}}_{F}={\pi\hbar^2\over{}N_Fm_e}\left[ZN_+d+{\pi\over2d^2}
 {N_F(N_F+1)(2N_F+1)\over6}\right],
\label{eq:42}
\end{equation}
where $N_+$ is the number of positive ions per unit volume, $Z$ is the
valence of these ions, and $N_F$ is the quantum index of the highest
occupied level. From Eq.~(\ref{eq:42}), the number of occupied levels
can be calculated if the thickness is known, and vice versa. The
minimal thickness for the quantum well to have $n$ levels below the
Fermi level can be determined from the relation
${\cal{E}}_{F}=\varepsilon_n$, and the maximal thickness from the
condition ${\cal{E}}_{F}=\varepsilon_{n+1}$. Thus for $n$ bound states
below the Fermi energy we find the minimal and maximal thicknesses
\begin{equation}
 d_{\rm{min}}^{n}=d_{\rm{max}}^{n-1}=
 \sqrt[3]{{\pi{}n\over2ZN_+}\left[n^2-{(n+1)(2n+1)\over6}\right]},
\label{eq:43}
\end{equation}
i.e., a result that depends on the number of levels below the Fermi
energy and the number of conduction electrons in the film.

To estimate the amount of phase conjugated light, we use the phase
conjugation (energy) reflection coefficient defined as
\begin{equation}
 R_{\rm{PC}}(\vec{q}_{\|},\omega)={I_{\rm{PC}}(-d;\vec{q}_{\|},\omega)
  \over{}I^{(1)}I^{(2)}I_{\rm{probe}}(-d;\vec{q}_{\|},\omega)},
\label{eq:44}
\end{equation}
in which $I^{(1)}$, $I^{(2)}$, $I_{\rm{probe}}$, and $I_{\rm{PC}}$ are
the intensities of the two pump beams, the probe and the phase
conjugated field, respectively. Each of the intensities are given by
$I={1\over2}{\epsilon_0c_0}{\vec{E}\cdot\vec{E}^{*}(2\pi)^{-4}}$. The
factor of $(2\pi)^{-4}$ originates from the manner in which we have
introduced the Fourier amplitudes of the fields.

For the remaining part of this work we choose a copper quantum well
with $N_+=8.47\times10^{28}$m$^{-3}$ and $Z=1$ (data taken from
Ref.~\onlinecite{Ashcroft:76:1}). The Cu quantum well is assumed to be
deposited on a glass substrate for which we use a refractive index $n$
of 1.51. With this substrate, the linear vaccum/substrate amplitude
reflection coefficients can be obtained by use of the classical
Fresnel formulae $r^s=[q_{\perp}-(n^2q^2-q_{\|}^2)^{1/2}]/
[q_{\perp}+(n^2q^2-q_{\|}^2)^{1/2}]$ and $r^p=[n^2q_{\perp}
-(n^2q^2-q_{\|}^2)^{1/2}]/[n^2q_{\perp}+(n^2q^2-q_{\|}^2)^{1/2}]$.
Having the pump wavevectors parallel to the $x$-axis then gives a pump
wavenumber of $k_{\|}=nq=1.51q$.

\section{Numerical results for a two-level quantum well}\label{sec:IV}
To calculate the phase conjugated response from a quantum well with an
arbitrary number of bound eigenstates one would have to superimpose
interband and intraband contributions. Thus in a study of the complete
response where local-field effects are neglected one basically would
have to add the contributions from the various pairs of levels located
in different subbands or in the same band. Seen in this light,
thorough treatments of the single-level case, where only intraband
transitions are allowed, and the two-level case, where transitions
between two eigenstates located in different bands occur, would form a
good qualitative starting point for analyses of multi-level
quantum-well systems. The single-level case we have studied
before,\cite{Andersen:98:2} and the following treatment will thus be
directed towards a description of the phase conjugated response from a
two-level quantum well. Thus, we choose the simplest possible
configuration in which interband transitions can occur, i.e., a
quantum well with only one bound state below the Fermi energy. Above
the Fermi energy we also assume that only one bound state can be
reached, and thus the wavefunctions are
$\psi_1(z)=\sqrt{2/d}\sin(\pi{}z/d)$ and
$\psi_2(z)=\sqrt{2/d}\sin(2\pi{}z/d)$. The associated energies then
become $\varepsilon_1=(\pi\hbar)^2/(2m_ed^2)$, and
$\varepsilon_2=(2\pi\hbar)^2/(2m_ed^2)$, respectively. The quantum
well with the various relevant energies and wavefunctions, as well as
the electronic excitations are shown in schematic form in
Fig.~\ref{fig:2}. In the present two-level case Eq.~(\ref{eq:40})
becomes
\begin{equation}
 p_{z,nm}={16\over3d}{\mathrm{sgn}\,}{(n-m)},
\label{eq:47}
\end{equation}
where $(n,m)\in\{(1,2),(2,1)\}$. If just the ground state should have
an energy less than the Fermi energy, we see from Eq.~(\ref{eq:43})
that the film thickness must be less than
$d_{\rm{max}}=\sqrt[3]{3\pi/(2ZN_+)}$. The minimal thickness is in the
IB model zero, but in reality the smallest thickness is a single
monolayer. Using Eq.~(\ref{eq:43}) the maximal thickness for a
two-level Cu quantum well then becomes
$d_{\rm{max}}\approx{}3.82${\AA}, which is more than two monolayers
and less than three. Thus we have two obvious choices for the
thickness of the quantum well, namely a single monolayer or two
monolayers. We choose two monolayers, since by this choice the two
energies $\varepsilon_1$ and $\varepsilon_2$ are closest to each
other, and thus the energy needed for a resonant transition to occur
is lowest. Two monolayers of copper roughly corresponds to a thickness
of $d=3.6$ {\AA} (bulk value). With this choice, the energy difference
between the two states is $\varepsilon_2-\varepsilon_1=8.70$ eV, and
the corresponding resonance in the optical spectrum is found at the
wavelength $\lambda=142.4$ nm.

\subsection{Phase conjugation reflection coefficient}
Among the eight possible ways of using linearly polarized light in our
chosen scattering configuration, two combinations give an
s-polarized response when using an s-polarized probe field, the
pump fields being either s-polarized or p-polarized, but with the
same polarization for both pump fields. When the pump fields are
s-polarized, the nonlinear conductivity tensor element that
contributes to the response is $\Xi_{yyyy}$. Altogether the phase
conjugated response in this purely s-polarized case (called ``sss'')
is negligible, since it is tens of orders of magnitude less than those
of the other combinations.  If, on the other hand, the pump fields are
p-polarized (pps), $\Xi_{yyzz}$ is the element of the nonlinear
conductivity tensor that contributes. Plotted as isophotes (contours
of equal intensity) in the normalized $\omega$-$q_{\|}$-plane
($\omega$ normalized to the interband transition frequency
$\omega_{21}$ and $q_{\|}$ normalized to the vacuum wavenumber
$\omega/c_0$), the result is shown in Fig.~\ref{fig:3}.

Two other combinations of polarization give p-polarized response using
a p-polarized probe field. As above, the pump fields have to be of the
same polarization, and can either be s- or p-polarized.  With
s-polarized pump fields (ssp), four tensor elements of the nonlinear
conductivity tensor contribute to the phase conjugated response (see
Table~\ref{tab:I}). The phase conjugated response is shown in the
normalized $\omega$-$q_{\|}$-plane in Fig.~\ref{fig:4}. In the other
case, another four tensor elements of the nonlinear conductivity
tensor contribute to the phase conjugated response when the pump
fields are p-polarized (see Table~\ref{tab:I}). We have in
Fig.~\ref{fig:5} shown the phase conjugated response for this
configuration (ppp) in the normalized $\omega$-$q_{\|}$-plane.

In the remaining four cases, the response has a different polarization
than the probe field. This is obtainable by the use of differently
polarized pump fields. In order to achieve an s-polarized response
from a p-polarized probe field one makes use of two differently
polarized pump fields, and four tensor elements of the DFWM response
tensor contribute to the solution, cf. Table~\ref{tab:I}. Similarly,
two differently polarized pump fields are needed in order to produce a
p-polarized response from an s-polarized source. For this process,
another four tensor elements of the nonlinear conductivity tensor
contributes according to Table~\ref{tab:I}. Since the resonance
structure of these last four cases are similar, it is sufficient here
to discuss the result obtained for just one of those cases. Thus, in
Fig.~\ref{fig:6} the result is shown for the case where pump field $1$
is s-polarized and pump field $2$ and the probe are p-polarized (spp).

\begin{intextfigure}
\begin{center}
\epsfig{file=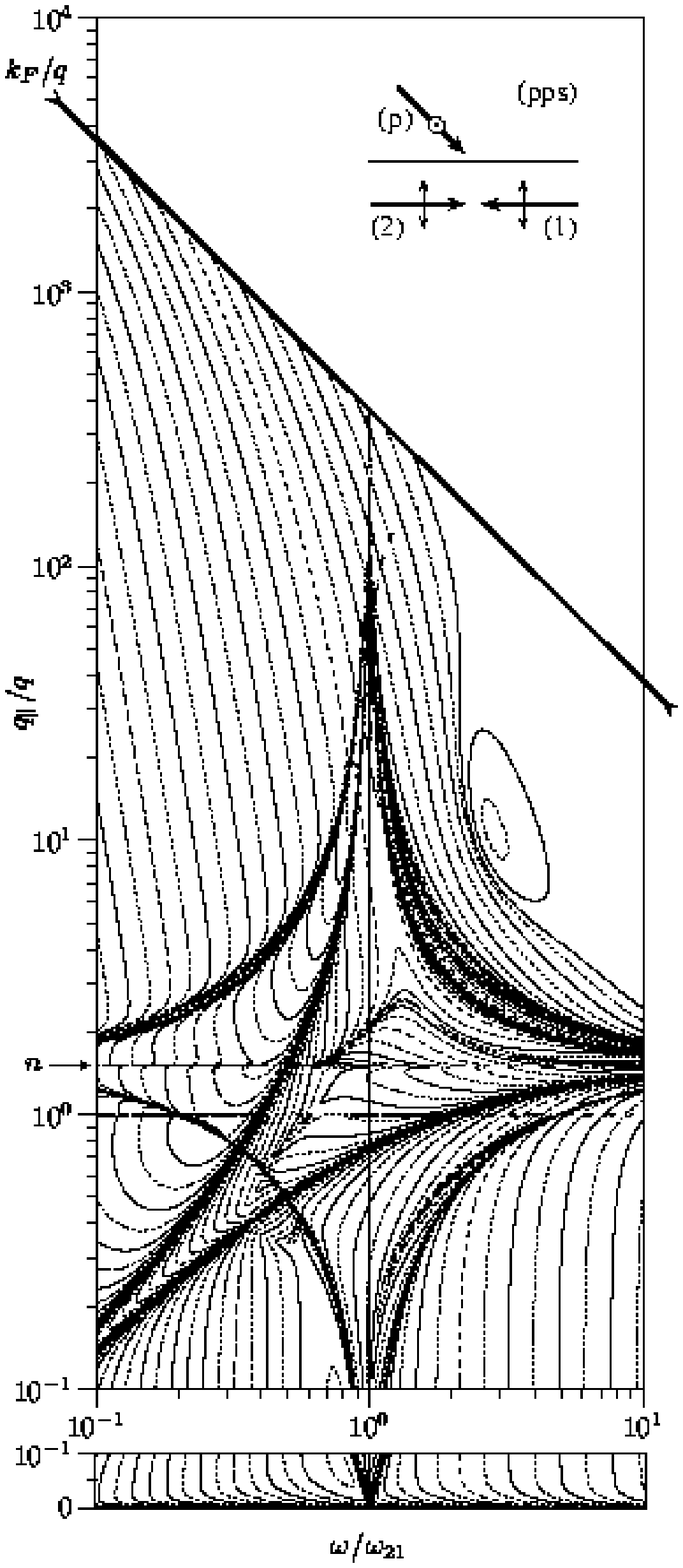,width=81mm}
\end{center}
\fcaption{The phase conjugation reflection coefficient from a two-level
  metallic quantum well is plotted in the case where s-polarized probe
  field gives s-polarized phase conjugated response, and where the
  pump fields are p-polarized (pps). The response is plotted as
  isophotes (contours of equal intensity) [m$^4$/W$^2$] on a
  logarithmic scale as a function of (i) the frequency $\omega$
  normalized to the transition frequency $\omega_{21}$, and (ii) the
  parallel component of the wavevector, normalized to the vacuum
  wavenumber. The difference between two neighbouring contours is one
  order of magnitude. To indicate the absolute amplitude, the isophote
  of value $10^{-20}$ m$^4$/W$^2$ has been plotted using a long-dashed
  curve and the isophote with magnitude $10^{-30}$ m$^4$/W$^2$ with a
  short-dashed curve. On the $q_{\|}/q$-scale, the response has been
  plotted on a linear scale in the range $0\leq{}q_{\|}/q\leq0.1$ and
  on a logarithmic scale above $q_{\|}/q=0.1$. \label{fig:3}}
\end{intextfigure}

\vskip -2\baselineskip
\begin{intextfigure}
\begin{center}
\epsfig{file=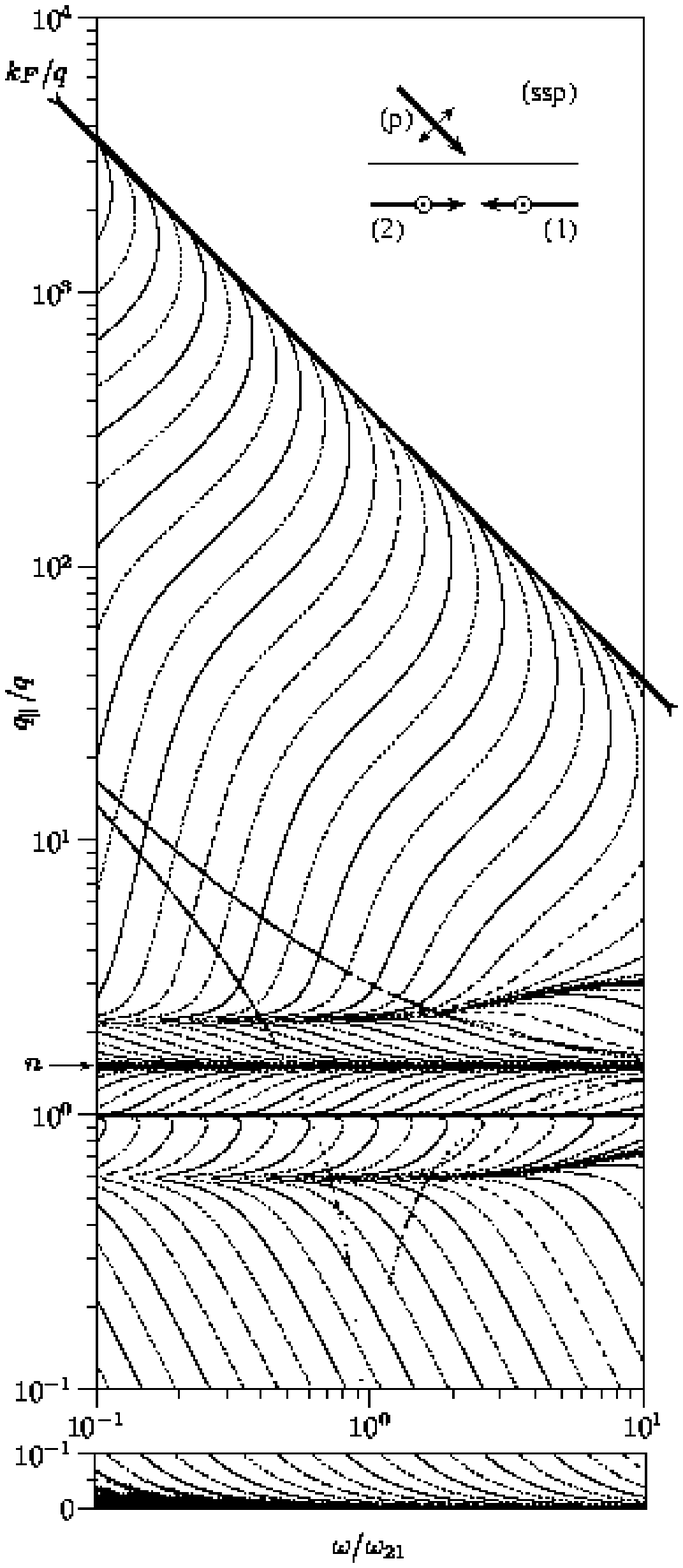,width=81mm}
\end{center}
\fcaption{The phase conjugation reflection coefficient from a two-level
  metallic quantum well is plotted in one of the cases where
  $p$-polarized probe field gives $p$-polarized phase conjugated
  response. In this case the pump fields are $s$-polarized (thus named
  ``ssp''). The response is plotted as isophotes [m$^4$/W$^2$] on a
  logarithmic scale as a function of (i) the frequency $\omega$
  normalized to the transition frequency $\omega_{12}$, and (ii) the
  parallel component of the wavevector, normalized to the vacuum
  wavenumber. The difference between two neighbouring isophotes is one
  order of magnitude. Again, the two isophotes of magnitude
  $10^{-20}$ m$^4$/W$^2$ and $10^{-30}$ m$^4$/W$^2$ has been plotted
  with long- and short-dashed curves, respectively. As before, below
  $0.1$, $q_{\|}/q$ has been plotted on a linear scale while above it
  is logarithmic. \label{fig:4}}
\end{intextfigure}

\begin{intextfigure}
\begin{center}
\epsfig{file=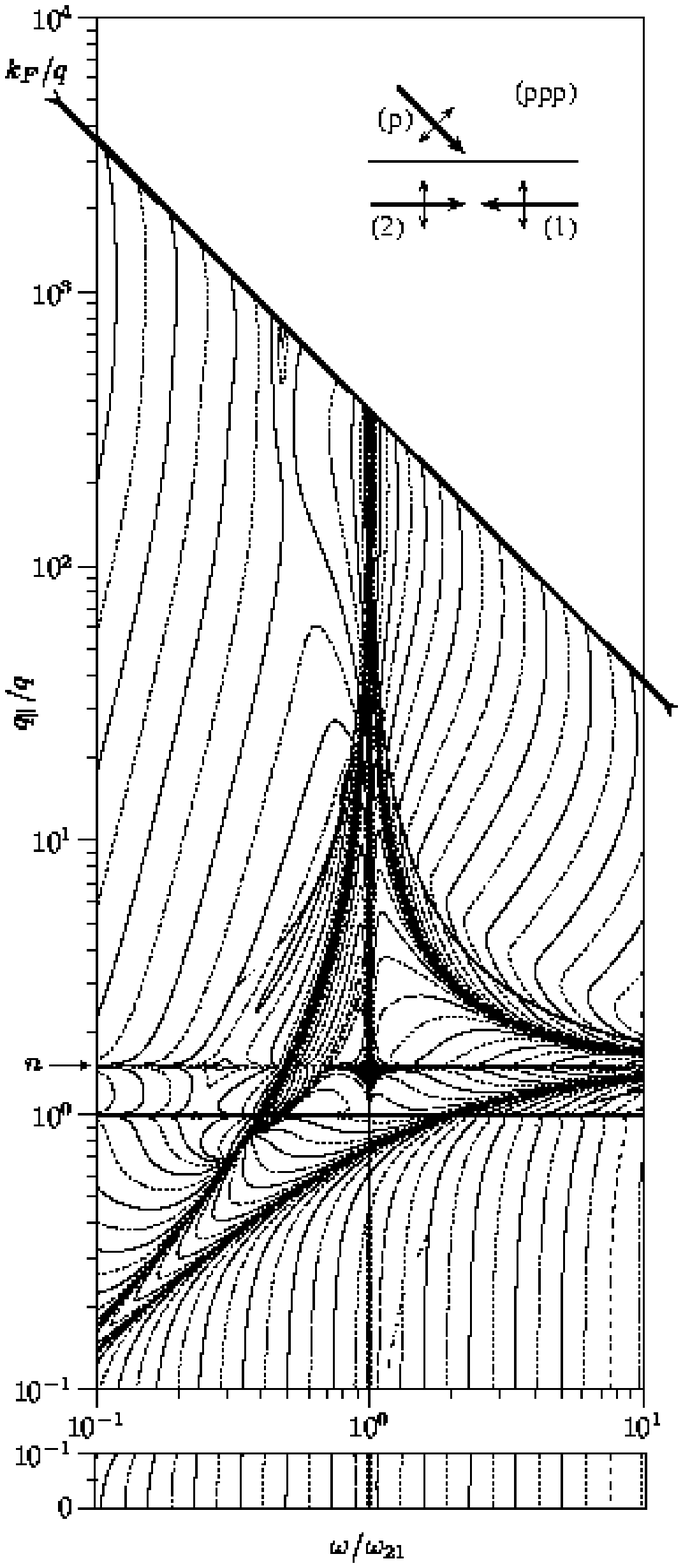,width=82mm}
\end{center}
\fcaption{The phase conjugation reflection coefficient from a two-level
  metallic quantum well is plotted in the other case where
  $p$-polarized probe field gives $p$-polarized phase conjugated
  response, this time with $p$-polarized pump fields (ppp). As in
  Figs.~\ref{fig:4} and \ref{fig:5}, the response is plotted as
  isophotes [m$^4$/W$^2$] on a logarithmic scale as a function of (i)
  the frequency $\omega$ normalized to the transition frequency
  $\omega_{12}$, and (ii) the parallel component of the wavevector,
  normalized to the vacuum wavenumber. Again, the difference between
  two neighbouring contours is one order of magnitude, and as before,
  the long- and short-dashed curves represents magnitudes of
  $10^{-20}$ m$^4$/W$^2$ and $10^{-30}$ m$^4$/W$^2$, respectively. In
  the big picture, $q_{\|}/q$ is plotted on a logarithmic scale, while
  in the strip it is plotted on a linear scale.\label{fig:5}}
\end{intextfigure}

\begin{intextfigure}
\begin{center}
\epsfig{file=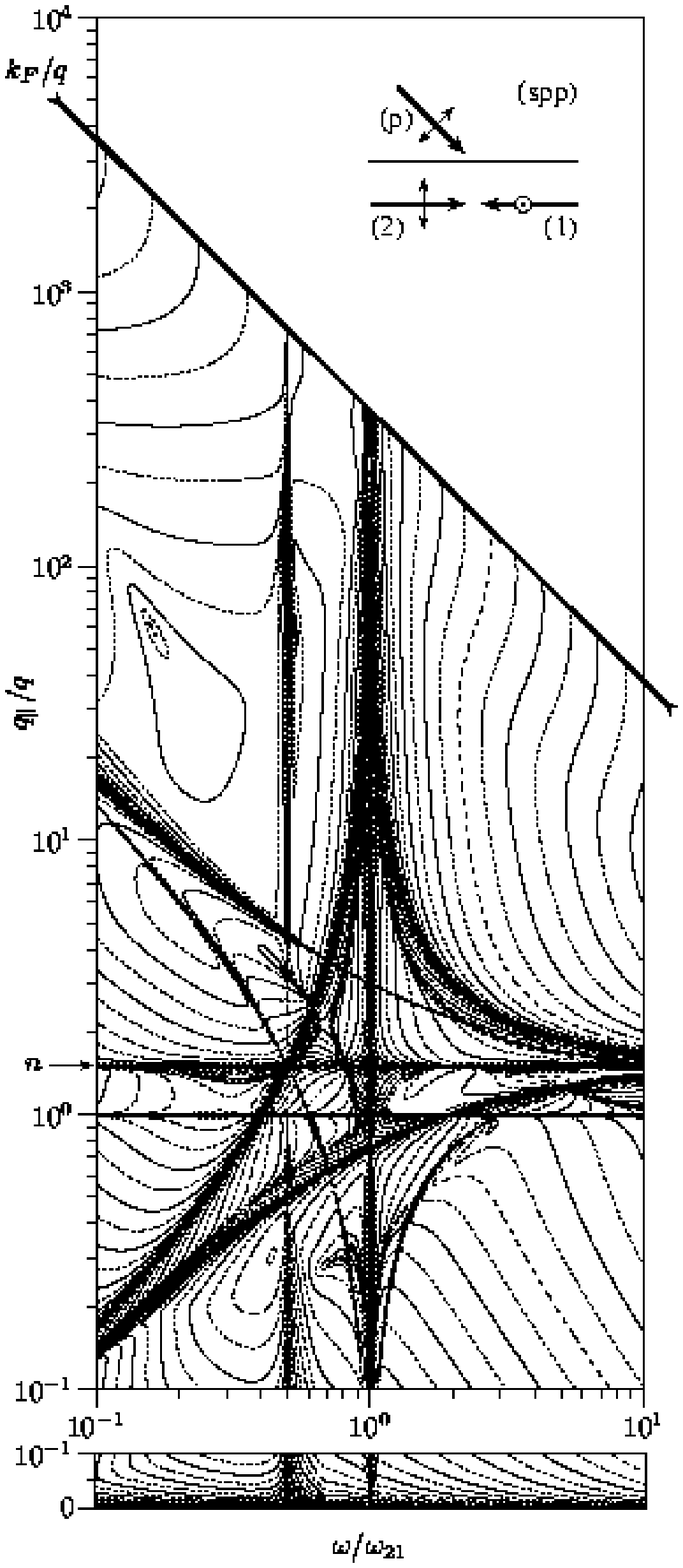,width=82mm}
\end{center}
\fcaption{The phase conjugation reflection coefficient from a two-level
  metallic quantum well is plotted in one of the cases where
  $p$-polarized probe field gives $s$-polarized phase conjugated
  response. In this case, pump field $1$ is $s$-polarized while pump
  field $2$ is $p$-polarized (spp). The response is plotted as
  isophotes [m$^4$/W$^2$] on a logarithmic scale as a function of (i)
  the frequency $\omega$ normalized to the transition frequency
  $\omega_{12}$, and (ii) the parallel component of the wavevector,
  normalized to the vacuum wavenumber. The difference between two
  neighbouring contours is one order of magnitude. The absolute
  amplitude of the isophote of value $10^{-20}$ m$^4$/W$^2$ has been
  plotted using a long-dashed curve and the isophote with magnitude
  $10^{-30}$ m$^4$/W$^2$ with a short-dashed curve. The strip below is
  plotted in a linear scale in $q_{\|}/q$ while the rest is on a
  logarithmic scale.\label{fig:6}}
\end{intextfigure}

The IB model only offers a crude description of the electronic
properties of a quantum well, since, for example, the electron density
profile at the ion/vacuum edge is poorly accounted for. This gives too
sharp a profile and underestimates the spill-out of the wavefunction.
Altogether one should be careful to put too much reality into the IB
model when treating local-field variations (related to, say, $q_{\|}$
or $q_{\perp}$) on the atomic length scale. Furthermore, neclecting
the Bloch character of the wavefunctions accounting for the dynamics
in the plane of the well is doubtful in investigations of the local
field among the atoms of the quantum well. The crucial quantity in the
above-mentioned context is the Fermi wave number
$k_F=(2m_e{\cal{E}}_{F})^{1/2}/\hbar$, and in relation to
Figs.~\ref{fig:3}--\ref{fig:6}, only results for $q_{\|}/q$ ratios
less than approximately
\begin{equation}
 {k_F\over{}q}=\lambda\sqrt{{ZN_+d\over2\pi}+{1\over4d^2}},
\label{eq:48}
\end{equation}
appears reliable. Thus we have cut off our results at the line
$q_{\|}/q=k_F/q$ in the $\omega/\omega_{21}$-$q_{\|}/q$-plane in
Figs.~\ref{fig:3}--\ref{fig:6}.

In many theoretical studies of the properties of phase conjugated
fields it is assumed that the phase conjugator is
ideal.\cite{Hendriks:89:1,Agarwal:95:1,Keller:96:2} By this is meant
that the phase conjugation reflection coefficient is independent of
the angle of incidense of the (propagating) probe field (and maybe
also of the state of polarization). As we concluded for the
single-level quantum well,\cite{Andersen:98:2} and as we can now see
for the two-level quantum well in Figs.~\ref{fig:3}--\ref{fig:6} this
assumption is not such a good approximation, at least not for a
metallic quantum well system.

\subsection{Resonant structure of the DFWM reflection coefficient} 
Looking at Figs.~\ref{fig:3}--\ref{fig:6}, a number of resonances
occur. They can all be accounted for from the analytic solution to
Eq.~(\ref{eq:A5}) by looking at the denominators appearing in the
analytic decomposition of the products, as given by Eqs.~(\ref{eq:B3})
and (\ref{eq:B4}) in Appendix~\ref{app:B}.  These resonances are shown
on the scale of Figs.~\ref{fig:3}--\ref{fig:6} in Fig.~\ref{fig:7}. In
the analytic solution of the integrals over $\vec{\kappa}_{\|}$ shown
in Appendix~\ref{app:B}, the solution to the terms with three
multiplied denominators is reduced in Eq.~(\ref{eq:B4}) to the problem
of finding a basic solution to the integrals over $\vec{\kappa}_{\|}$
for each of these denominators multiplied by a
$\vec{\kappa}_{\|}$-independent factor. The resulting integrals do not
contain sharp resonances, but the factors in front of them do, when
$a_ib_j-b_ia_j=0$, for $i,j\in\{1,2,3\}$ and $i\neq{}j$. In order to
make an analytical treatment of the resonances appearing in the
nonlinear conductivity tensor we in the following define a term of the
nonlinear conductivity tensor as a product of three denominators in
Eq.~(\ref{eq:A5}), and number them $1,2,\dots,12$.  However, not all
terms gives contributions to the result in a two-level quantum well.
The terms that does not give any contributions are the terms with a
$2\omega$-contribution in the denominator, i.e., terms 1--2 and
11--12. When the denominators of the rest of the terms (3--10) are put
into the form of Eq.~(\ref{eq:B1}), a total of four different $a$'s
and nine different $b$'s appear. They are listed in
Appendix~\ref{app:C}. Since we are looking for the location of the
resonances in the system it is reasonable in the following analysis
to let the respective relaxation times $\tau_{nm}$ in
Eqs.~(\ref{eq:C5})--(\ref{eq:C13}) be infinite.

\begin{intextfigure}
\setlength{\unitlength}{1.1mm}
\psset{unit=1.1mm}
\begin{pspicture}(0,-2)(75,163)
\put(0,0){
 \put(10,10){\scalebox{-1}[1]{%
  \rput[bl]{90}(0,0){\epsfig{file=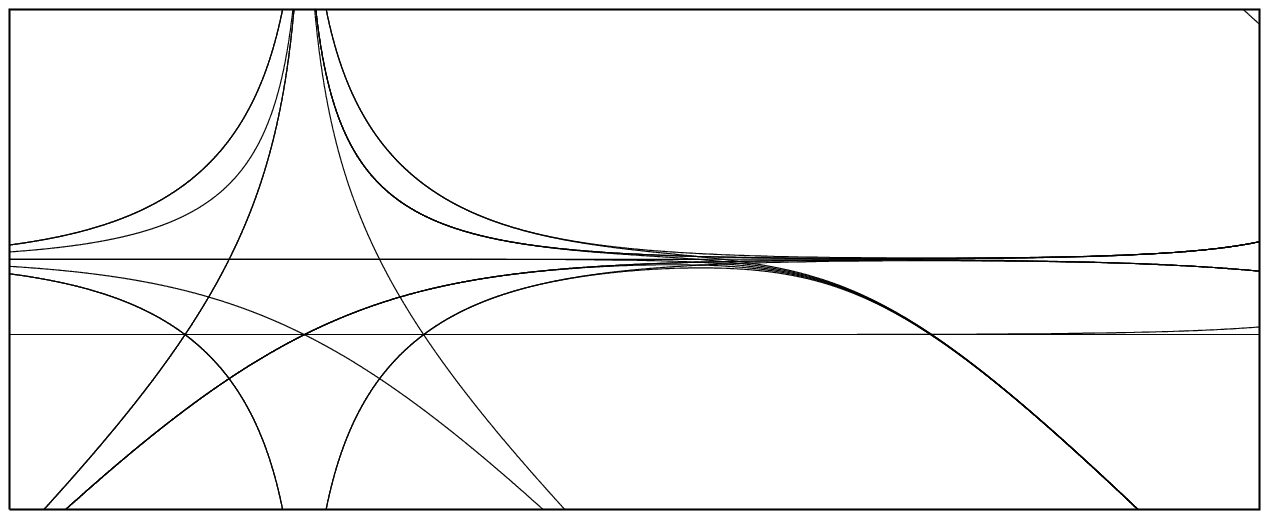,width=150\unitlength}}%
 }}
 \put(10,10){%
  \setlength{\decade}{30\unitlength}
  \psline[linewidth=0.15mm]{-}(0.699\decade,0)(0.699\decade,150)
 }
 \psline[linewidth=0.15mm]{-}(10,45.369)(70,45.369)
 \setcounter{puppet}{-1}
 \multiput(10,10)(30,0){3}{
  \psline[linewidth=0.25mm]{-}(0,0)(0,-2)
  \put(0,-2.5){\makebox(0,0)[t]{$10^{\arabic{puppet}}$}}
  \addtocounter{puppet}{1}
 }
 \multiput(10,10)(30,0){2}{
  \setlength{\decade}{30\unitlength}
  \psline[linewidth=0.25mm]{-}(0.301\decade,0)(0.301\decade,-1)
  \psline[linewidth=0.25mm]{-}(0.477\decade,0)(0.477\decade,-1)
  \psline[linewidth=0.25mm]{-}(0.602\decade,0)(0.602\decade,-1)
  \psline[linewidth=0.25mm]{-}(0.699\decade,0)(0.699\decade,-1)
  \psline[linewidth=0.25mm]{-}(0.778\decade,0)(0.778\decade,-1)
  \psline[linewidth=0.25mm]{-}(0.845\decade,0)(0.845\decade,-1)
  \psline[linewidth=0.25mm]{-}(0.903\decade,0)(0.903\decade,-1)
  \psline[linewidth=0.25mm]{-}(0.954\decade,0)(0.954\decade,-1)
 }
 \setcounter{puppet}{-1}
 \multiput(10,10)(0,30){6}{
  \psline[linewidth=0.25mm]{-}(0,0)(-2,0)
  \put(-2.5,0){\makebox(0,0)[r]{$10^{\arabic{puppet}}$}}
  \addtocounter{puppet}{1}
 }
 \multiput(10,10)(0,30){5}{
  \setlength{\decade}{30\unitlength}
  \psline[linewidth=0.25mm]{-}(0,0.301\decade)(-1,0.301\decade)
  \psline[linewidth=0.25mm]{-}(0,0.477\decade)(-1,0.477\decade)
  \psline[linewidth=0.25mm]{-}(0,0.602\decade)(-1,0.602\decade)
  \psline[linewidth=0.25mm]{-}(0,0.699\decade)(-1,0.699\decade)
  \psline[linewidth=0.25mm]{-}(0,0.778\decade)(-1,0.778\decade)
  \psline[linewidth=0.25mm]{-}(0,0.845\decade)(-1,0.845\decade)
  \psline[linewidth=0.25mm]{-}(0,0.903\decade)(-1,0.903\decade)
  \psline[linewidth=0.25mm]{-}(0,0.954\decade)(-1,0.954\decade)
 }
 \put(40,0){\makebox(0,0)[b]{$\omega/\omega_{21}$}}
 \rput[t]{90}(0,85){\makebox(0,0)[t]{$q_{\|}/q$}}
 \pspolygon[linewidth=0mm,linecolor=white,fillstyle=solid,fillcolor=white](10,147)(70,87)(70,160)(10,160)
 \psline[linewidth=0.20mm]{-}(10,10)(70,10)(70,160)(10,160)(10,10)
 \psline[linewidth=0.5mm]{>-<}(5,151.5)(73.5,84)
 \put(0,154){\makebox(0,0)[l]{$k_{F}/q$}}
 \psline[linewidth=0.25mm]{->}(5,45.369)(9.5,45.369)
 \put(4,45.369){\makebox(0,0)[r]{$n$}}
 \put(17,20){\makebox(0,0)[tl]{a}}
 \put(15,23){\makebox(0,0)[br]{b}}
 \put(19.5,62){\makebox(0,0)[tr]{c}}
 \put(22,67){\makebox(0,0)[bl]{d}}
 \put(48.5,56){\makebox(0,0)[bl]{e}}
 \put(46.5,33){\makebox(0,0)[tl]{f}}
 \put(30,75){\makebox(0,0)[br]{g}}
 \put(40.5,60){\makebox(0,0)[bl]{h}}
 \put(51,58){\makebox(0,0)[bl]{i}}
 \put(49,30.5){\makebox(0,0)[tl]{j}}
 \put(21,40){\makebox(0,0)[bl]{k}}
 \put(19,52){\makebox(0,0)[bl]{l}}
 \put(27,125.5){\makebox(0,0)[tr]{m}}
}
\end{pspicture}
\fcaption{Resonances of the nonlinear conductivity tensor are shown as
  a function of (i) the optical frequency normalized to the transition
  frequency ($\omega/\omega_{21}$) and (ii) the parallel component of
  the wavevector normalized to the vacuum wavenumber ($q_{\|}/q$).
  This figure shows only the pure resonances. The broadening due to
  the relaxation times is neglected by setting them all to
  infinity.\label{fig:7}}
\end{intextfigure}

In terms of the $a$'s and $b$'s listed in Appendix~\ref{app:C}, we
observe that the third term of $\Xi_{ijkh}$ has resonances at
(i) $a_1b_{nm}^4-b_{vl}^2a_2=0$, (ii) $a_3b_{nm}^4-b_{nl}^6a_2=0$, and
(iii) $a_3b_{vl}^2-b_{nl}^6a_1=0$. After insertion of the relevant
$a$'s and $b$'s, substitution of $k_{\|}$ in favor of $n\omega/c_0$
(since $k_{\|}=n\omega/c_0$ in our treatment), and a normalization of
$q_{\|}$ to $q$, i.e., $q_{\|}=(q_{\|}/q)\omega/c_0$, we may solve the
resulting second order equations with respect to $\omega$ as a
function of $q_{\|}/q$. Then resonance condition (i) gives
\endmulticols
\begin{eqnarray}
 \omega=
 {m_ec_0^2\over\hbar{}nq_{\|}/q}{n-q_{\|}/q\over{}n+q_{\|}/q}
 \pm\sqrt{
 \left({m_ec_0^2\over\hbar{}nq_{\|}/q}{n-q_{\|}/q\over{}n+q_{\|}/q}\right)^2
 +{2m_ec_0^2\over\hbar^2(n+q_{\|}/q)}
 \left[{\epsilon_v-\epsilon_l\over{}n}+{\epsilon_m-\epsilon_n\over{}q_{\|}/q}
 \right]}
,\label{eq:49}
\end{eqnarray}
resonance condition (ii) becomes
\begin{eqnarray}
 \omega=
 {m_ec_0^2\over\hbar{}nq_{\|}/q}
 \pm\sqrt{\left({m_ec_0^2\over\hbar{}nq_{\|}/q}\right)^2
 +{2m_ec_0^2\over\hbar^2n}\left[{\epsilon_n-\epsilon_l\over{}n+q_{\|}/q}
 +{\epsilon_m-\epsilon_n\over{}q_{\|}/q}\right]}
,\label{eq:50}
\end{eqnarray}
and condition (iii) is
\begin{eqnarray}
 \omega=
 -{m_ec_0^2\over\hbar{}nq_{\|}/q}
 \pm\sqrt{\left({m_ec_0^2\over\hbar{}nq_{\|}/q}\right)^2
 +{2m_ec_0^2\over\hbar^2q_{\|}/q}\left[{\epsilon_v-\epsilon_l\over{}n}
 +{\epsilon_l-\epsilon_n\over{}n+q_{\|}/q}\right]}
.\label{eq:51}
\end{eqnarray}
\narrowtext\noindent
In some of the above equations, some of the solutions can be ruled out
immediately, since, for example, in Eq.~(\ref{eq:50}) the minus in
front of the square root gives only rise to negative values of
$\omega$ in the ``interchanged term'' (when $k_{\|}=-n\omega/c_0$).

In the fourth term of $\Xi_{ijkh}$ we observe that in
addition to a resonance of type (ii), resonances appear at (iv)
$a_2b_{nm}^4-b_{nv}^3a_2=0$ and (v) $a_3b_{nv}^3-b_{nl}^6a_2=0$. Again
inserting the respective $a$'s and $b$'s from Appendix~\ref{app:C},
substituting $n\omega/c_0$ for $k_{\|}$, and normalizing $q_{\|}$ to
the vacuum wavenumber, resonance condition (iv) becomes
\begin{eqnarray}
 \omega={1\over2\hbar}(\epsilon_v-\epsilon_m)
,\label{eq:52}
\end{eqnarray}
and resonance condition (v) is equivalent to Eq.~(\ref{eq:50}), taking
into account the interchanged term. In our configuration, the choice
of a two-level quantum well puts some restrictions on the values of
the quantum numbers $n$, $m$, $v$, and $l$ in order to get a nonzero
result. Comparing Eqs.~(\ref{eq:A5}) and (\ref{eq:38}) we observe that
if pump field one (indexed $k$) is s-polarized then $l=v$, while
$l\neq{}v$ if it is p-polarized.  Similarly, if the other pump field
(indexed $h$) is s-polarized we get $m=l$, while we get $m\neq{}l$ if
it is p-polarized. These conditions are summarized in
Table~\ref{tab:II}, and the contributions from
Eqs.~(\ref{eq:49})--(\ref{eq:52}) to the resonances in
Fig.~\ref{fig:7} are shown in Table~\ref{tab:III} for the valid
combinations of quantum numbers.

\begin{intexttable}
\tcaption{Restrictions on the valid combinations of quantum numbers for
  a two-level quantum well in the nonlinear conductivity tensor
  for the three combinations of polarized light of the pump fields
  treated in this communication. Pump field 1 is indexed $k$, and pump
  field 2 $h$ in Eq.~(\ref{eq:A5}).\label{tab:II}}
\noindent
\begin{tabular*}{8.6cm}{cc|@{\extracolsep{\fill}}ccc}
\tableline\tableline
 $k$ & $h$ & $\Xi$ terms 3--4 & $\Xi$ terms 5--8 & $\Xi$ terms 9--10 \\
 \tableline
 s & s & $l=v=m$ & $v=n\wedge{}m=l$ & $v=n=l$ \\
 s & p & $l=v\wedge{}m\neq{}l$ & $v=n\wedge{}m\neq{}l$ &$v=n\wedge{}l\neq{}v$\\
 p & p & $m\neq{}l\wedge{}l\neq{}v$ & $v\neq{}n\wedge{}m\neq{}l$
 &$v\neq{}n\wedge{}l\neq{}v$ \\
\tableline\tableline
\end{tabular*}
\end{intexttable}
The resonances conditions in the fifth term of $\Xi_{ijkh}$
are (vi) $a_1b_{nm}^4+b_{lm}^1a_2=0$, (vii)
$a_4b_{nm}^4-b_{vm}^5a_2=0$, and (viii) $a_4b_{lm}^1+b_{vm}^5a_1=0$.
By insertion of the respective $a$'s and $b$'s from
Appendix~\ref{app:C}, substitution of $k_{\|}$ by $n\omega/c_0$, and
normalization of $q_{\|}$ to the vacuum wavenumber, resonance
condition (vi) becomes 
\widetext
\begin{eqnarray}
 \omega&=&
 -{m_ec_0^2\over\hbar{}(q_{\|}/q)^2}
 \pm\sqrt{\left({m_ec_0^2\over\hbar{}(q_{\|}/q)^2}\right)^2
 +{2m_ec_0^2n\over\hbar^2(n-q_{\|}/q)q_{\|}/q}\left[
 {\epsilon_n-\epsilon_m\over{}q_{\|}/q}+{\epsilon_v-\epsilon_l\over{}n}\right]}
,\label{eq:54}
\end{eqnarray}
condition (vii) gives
\begin{eqnarray}
 \omega&=&
 {m_ec_0^2\over\hbar{}nq_{\|}/q}
 \pm\sqrt{\left({m_ec_0^2\over\hbar{}nq_{\|}/q}\right)^2
 +{2m_ec_0^2\over\hbar^2n}\left[{\epsilon_m-\epsilon_n\over{}q_{\|}/q}
 +{\epsilon_m-\epsilon_v\over{}n-q_{\|}/q}\right]},
\label{eq:55}
\end{eqnarray}
and case (viii) becomes
\begin{eqnarray}
 \omega&=&
 {m_ec_0^2\over\hbar{}nq_{\|}/q}
 \pm\sqrt{\left({m_ec_0^2\over\hbar{}nq_{\|}/q}\right)^2
 +{2m_ec_0^2\over\hbar^2q_{\|}/q}\left[{\epsilon_m-\epsilon_l\over{}n}
 +{\epsilon_v-\epsilon_m\over{}n-q_{\|}/q}\right]}.
\label{eq:56}
\end{eqnarray}
The sixth term of $\Xi_{ijkh}$ has a resonance of the type
(vi), and further resonances at (ix) $a_2b_{nm}^4-b_{vl}^7a_2=0$ and
(x) $a_4b_{vl}^7-b_{vm}^5a_2=0$. Insertion of the different $a$'s and
$b$'s, $k_{\|}=n\omega/c_0$, and nomalizing $q_{\|}$ to the the vacuum
wavenumber gives (ix) resonances at
\begin{eqnarray}
 \omega&=&
 {m_ec_0^2\over\hbar{}nq_{\|}/q}
 \pm\sqrt{\left({m_ec_0^2\over\hbar{}nq_{\|}/q}\right)^2
 +{m_ec_0^2\over\hbar^2nq_{\|}/q}
 \left[\epsilon_m+\epsilon_v-\epsilon_n-\epsilon_l\right]},
\label{eq:57}
\end{eqnarray}
and (x) resonances equivalent to those given in Eq.~(\ref{eq:55}). In
the seventh term of $\Xi_{ijkh}$ there is a resonances of the type of
case (ix), and furthermore at (xi) $a_3b_{nm}^4-b_{nl}^6a_2=0$ and
(xii) $a_3b_{vl}^7-b_{nl}^6a_2=0$. As in the previous cases we insert
the different $a$'s and $b$'s found in Appendix~\ref{app:C}, replace
$k_{\|}$ with $n\omega/c_0$, and nomalize $q_{\|}$ to the the vacuum
wavenumber. Then case (xi) gives resonances at
\begin{eqnarray}
 \omega=
 {m_ec_0^2\over\hbar{}nq_{\|}/q}{n+q_{\|}/q\over{}n-q_{\|}/q}
 \pm\sqrt{\left({m_ec_0^2\over\hbar{}nq_{\|}/q}{n+q_{\|}/q\over{}n-q_{\|}/q}
 \right)^2+{2m_ec_0^2\over\hbar^2n}\left[{n+q_{\|}/q\over{}n-q_{\|}/q}
 {\epsilon_n-\epsilon_m\over{}q_{\|}/q}
 +{\epsilon_l-\epsilon_n\over{}n-q_{\|}/q}\right]}
,\label{eq:59}
\end{eqnarray}
and case (xii) the resonances are equivalent to Eq.~(\ref{eq:50}).
The eighth term of $\Xi_{ijkh}$ has a resonance of the type given in
case (xi), and additional resonances at (xiii)
$a_1b_{nm}^4-b_{nv}^8a_2=0$ and (xiv) $a_3b_{nv}^8-b_{nl}^6a_1=0$.
Repeating the procedure from above, we get for case (xiii) the
solution
\begin{eqnarray}
 \omega&=&
 {m_ec_0^2\over\hbar{}nq_{\|}/q}
 \pm\sqrt{\left({m_ec_0^2\over\hbar{}nq_{\|}/q}\right)^2
 +{2m_ec_0^2\over\hbar^2(n-q_{\|}/q)}\left[
 {\epsilon_m-\epsilon_n\over{}q_{\|}/q}+{\epsilon_n-\epsilon_v\over{}n}\right]}
,\label{eq:61}
\end{eqnarray}
\narrowtext\noindent 
and in case (xiv) gives resonances equivalent to the result given in
Eq.~(\ref{eq:51}). Again, when considering a two-level quantum well in
our configuration, some restrictions apply to the quantum numbers. If
we again compare Eqs.~(\ref{eq:A5}) and (\ref{eq:38}) we see that if
pump field one (index $k$) is s-polarized, then $v=n$, and if it is
p-polarized, then $v\neq{}n$.  Additionally, if pump field two (index
$h$) is s-polarized, $m=l$, and if it is p-polarized, $m\neq{}l$.
This has the consequences that (i) the quantum numbers $n$ and $m$ can
be chosen arbitrarily when both pump fields are s-polarized, (ii) when
both pump fields are p-polarized we either get $m=n$ and $l=v$, or we
get $m=v$ and $l=n$, (iii) when pump field one is s-polarized and the
other one p-polarized we get either $m=v$ or $l=v$, and (iv) in the
opposite case we get either $m=v$ or $m=n$. These conditions are
summarized in Table~\ref{tab:II}, and the contributions from
Eqs.~(\ref{eq:54})--(\ref{eq:61}) to the resonances in
Fig.~\ref{fig:7} are shown in Table~\ref{tab:III} for the valid
combinations of quantum numbers. It should be noted that in
Eq.~(\ref{eq:54}), the combinations of quantum numbers that give rise
to the resonances ``b'', ``e'', ``i'', ``h'', and ``l'' are going into
resonance ``m'' after they have reached the line at
$\omega/\omega_{21}=1$.  None of the other equations contributes to
resonance ``m''.

For the ninth term of $\Xi_{ijkh}$ the resonances are at (xv)
$a_2b_{nm}^4-b_{lm}^3a_2=0$, (xvi) $a_4b_{nm}^4-b_{vm}^5a_2=0$, and
(xvii) $a_4b_{lm}^3-b_{vm}^5a_2=0$.  After insertion of the relevant
$a$'s and $b$'s from Eqs.~(\ref{eq:C1})--(\ref{eq:C13}),
$k_{\|}=n\omega/c_0$ and a normalization of $q_{\|}$ to the vacuum
wavenumber, we resulting second order equations can be solved with
respect to $\omega$ as a function of $q_{\|}/q$. Then case (xv) is
equivalent to Eq.~(\ref{eq:52}), and cases (xvi) and (xvii) to
Eq.~(\ref{eq:55}). Finally, in the tenth term of $\Xi_{ijkh}$ a
resonance of the type given by case (xvi) occur. Two other resonances
are located at (xviii) $a_1b_{nm}^4+b_{vl}^9a_2=0$ and at (xix)
$a_4b_{vl}^9+b_{vm}^5a_1=0$, respectively. Inserting the $a$'s and
$b$'s given in Appendix~\ref{app:C} and using the same substitution
and normalization as above, case (xviii) gives
\endmulticols
\begin{table}
\caption{Resonances generated by Eqs.~(\ref{eq:49})--(\ref{eq:66})
  are shown as a function of the valid combinations of quantum numbers
  $(n,m,v,l)$ and the sign appearing in front of the square roots. In
  each of upper and lower parts of the table, the upper row shows the
  generating equation and the next four rows show the values of the
  quantum numbers, which can take the value $1$ or $2$ in a two-level
  quantum well. The last four rows show the resonances resulting from
  use of the quantum numbers in the respective equations for each sign
  $+$ and $-$, the first two of these rows being associated with the
  normal term, and the last two with the interchanged term. A zero in
  the last four rows refers to $\omega=0$, and the letters a--l refers
  to the resonances shown in Fig.~\ref{fig:7}. A ``$*$'' is used when
  the value of a quantum number is indifferent, and a ``-'' in the
  output field appears when the result is outside the shown range in
  Fig.~\ref{fig:7}. Since Eq.~(\ref{eq:52}) is a linear solution in
  $\omega$ the sign does not apply, and the result is listed under the
  sign ``+'' for simplicity.  It should be noted that in
  Eq.~(\ref{eq:54}), that the combinations of quantum numbers that
  give rise to the resonances ``b'', ``e'', ``i'', ``h'', and ``l''
  are going into resonance ``m'' after they have reached the line at
  $\omega/\omega_{21}=1$.\label{tab:III}}
\begin{tabular}{c|ccccccc|ccccccc|ccccccc|ccc|ccccccccc|ccccccc}
 &\multicolumn{7}{c|}{Eq.~(\ref{eq:49})} &
  \multicolumn{7}{c|}{Eq.~(\ref{eq:50})} &
  \multicolumn{7}{c|}{Eq.~(\ref{eq:51})} &
  \multicolumn{3}{c|}{Eq.~(\ref{eq:52})}
 &\multicolumn{9}{c|}{Eq.~(\ref{eq:54})}
 &\multicolumn{7}{c}{Eq.~(\ref{eq:55})} \\
 \tableline
 $n$&$*$&1&1&1&2&2&2&$*$&1&1&1&2&2&2&$*$&1&1&1&2&2&2&$*$&$*$&$*$&$*$&$*$&$*$&1&1&1&2&2&2&$*$&1&1&1&2&2&2 \\
 $m$&$n$&1&2&2&1&1&2&$n$&1&2&2&1&1&2&$*$&$*$&$*$&$*$&$*$&$*$&$*$&$*$&1&2&$n$&$n$&$n$&2&2&2&1&1&1&$n$&1&2&2&1&1&2 \\
 $v$&$*$&1&$*$&2&$*$&1&2&$*$&$*$&$*$&$*$&$*$&$*$&$*$&$n$&1&2&2&1&1&2&$m$&2&1&$*$&1&2&1&1&2&1&2&2&$n$&2&1&2&1&2&1 \\
 $l$&$v$&2&$v$&1&$v$&2&1&$n$&2&1&2&1&2&1&$n$&2&1&2&1&2&1&$*$&$*$&$*$&$v$&2&1&1&2&1&2&1&2&$*$&$*$&$*$&$*$&$*$&$*$&$*$ \\
  \tableline 
 $+$&-&-&d&i&-&-&e&-&-&-&-&-&-&-&0&-&h&c&-&h&-&0&g&-&0&e&a&d&i&-&h&j&f&-&-&-&-&-&-&- \\
 $-$&0&a&-&-&-&j&-&0&b&-&-&c&h&-&-&-&-&-&-&-&-& & & &-&-&-&-&-&-&-&-&-&0&a&d&-&h&f&e  \\
  \tableline 
 $+$&0&-&-&-&c&k&b&0&e&-&d&f&h&a&-&-&-&-&-&-&-&0&g&-&0&b&-&-&l&-&h&k&c&0&-&-&-&h&c&b \\
 $-$&-&-&-&l&c&-&b&-&-&-&-&-&-&-&0&-&h&f&d&h&-& & & &-&-&-&-&-&-&-&-&-&-&-&-&-&-&-&-
\end{tabular}
\begin{tabular}{c|ccccccc|cccccccc|ccccccc|ccccccc|ccccccccc}
&\multicolumn{7}{c|}{Eq.~(\ref{eq:56})}
 &\multicolumn{8}{c|}{Eq.~(\ref{eq:57})}
 &\multicolumn{7}{c|}{Eq.~(\ref{eq:59})}
 &\multicolumn{7}{c|}{Eq.~(\ref{eq:61})}
 &\multicolumn{9}{c}{Eq.~(\ref{eq:66})}\\
 \tableline
 $n$&$*$&$*$&$*$&$*$&$*$&$*$&$*$&$*$&$*$&$*$&1&$*$&1&2&2&$*$&1&1&1&2&2&2&$*$&1&1&1&2&2&2&$*$&$*$&$*$&1&1&1&2&2&2 \\
 $m$&$*$&1&1&1&2&2&2&$n$&$n$&$n$&2&$*$&2&1&1&$n$&1&2&2&1&1&2&$n$&1&2&2&1&1&2&$n$&$n$&$n$&2&2&2&1&1&1 \\
 $v$&$m$&1&2&2&1&1&2&$*$&1&2&1&$n$&2&1&2&$*$&$*$&$*$&$*$&$*$&$*$&$*$&$n$&2&1&2&1&2&1&$*$&1&2&1&1&2&1&2&2 \\
 $l$&$m$&2&1&2&1&2&1&$v$&2&1&1&$m$&1&2&2&$n$&2&1&2&1&2&1&$*$&$*$&$*$&$*$&$*$&$*$&$*$&$v$&2&1&1&2&1&2&1&2 \\
 \tableline
 $+$&-&-&-&-&-&-&-&-&-&-&-&-&-&-&-&-&-&h&c&-&-&b&-&-&-&-&-&-&-&-&-&b&-&-&l&-&h&c \\
 $-$&0&h&d&e&a&f&-&0&g&-&-&0&-&h&g&0&-&h&c&-&-&b&0&a&d&-&h&f&e&0&-&b&-&-&-&k&h&c \\
 \tableline
 $+$&0&h&-&b&-&c&-&0&g&-&-&0&-&h&g&0&a&h&f&-&-&-&0&-&-&-&h&c&b&0&a&-&-&-&-&j&h&f \\
 $-$&-&-&-&-&-&-&-&-&-&-&-&-&-&-&-&-&-&h&-&d&-&e&-&-&-&-&-&-&-&-&-&e&d&-&i&-&h&- \\
\end{tabular}
\end{table}

\begin{eqnarray}
 \omega=
 {m_ec_0^2\over\hbar{}nq_{\|}/q}{n+q_{\|}/q\over{}n-q_{\|}/q}
 \pm\sqrt{\left({m_ec_0^2\over\hbar{}nq_{\|}/q}{n+q_{\|}/q\over{}n-q_{\|}/q}
 \right)^2+{2m_ec_0^2\over\hbar^2(n-q_{\|}/q)}\left[
 {\epsilon_m-\epsilon_n\over{}q_{\|}/q}+{\epsilon_l-\epsilon_v\over{}n}\right]}
,\label{eq:66}
\end{eqnarray}
\narrowtext\noindent 
and case (xix) has a solution equivalent to the one given in
Eq.~(\ref{eq:56}). As before we find by a comparison of
Eqs.~(\ref{eq:A5}) and (\ref{eq:38}) that some selection rules appear
when choosing a two-level quantum well in our configuration, since
when pump field one (indexed $k$) is s-polarized we get $v=n$, and
when it is p-polarized, $v\neq{}n$. Similarly, when pump field two
(indexed $h$) is s-polarized we get $l=v$, and when it is p-polarized,
$l\neq{}v$. Then, if both pump fields are s-polarized we may in a
two-level quantum well choose $m=n$ or $m\neq{}n$. In the case where
both pump fields are p-polarized, the result is identically zero. In
the case where pump field one is s-polarized and pump field two is
p-polarized we may choose either $m=l$ or $m=v$, while in the opposite
case we may choose either $m=l$ or $m=n$. As before, these conditions
are summarized in Table~\ref{tab:II}, and the contributions from
Eq.~(\ref{eq:66}) to the resonances in Fig.~\ref{fig:7} are shown in
Table~\ref{tab:III} for the valid combinations of quantum numbers.

In the linear conductivity tensor [Eq.~(\ref{eq:7})] resonances occur
when $a_2b_{nm}^4-a_2b_{nm}^{10}=0$, where
\begin{equation}
 b_{nm}^{10}={1\over\hbar}(\epsilon_n-\epsilon_m)
 +{\hbar{}q_{\|}^2\over2m_e}-i\tau_{nm}^{-1}
.\label{eq:68}
\end{equation}
The solutions are $q_{\|}=0$ or $\omega=0$, independent of the values
of $n$ and $m$. Adding this resonance to the ones we found in
Eqs.~(\ref{eq:49})--(\ref{eq:66}) the resonances appearing in
Figs.~\ref{fig:3}--\ref{fig:6} have been identified. Q.~E.~D.

While most of the resonances described above and shown in
Fig.~\ref{fig:7} are clearly pronounced in
Figs.~\ref{fig:3}--\ref{fig:6}, the resonance named ``m'' does not
appear so clearly, although in Figs.~\ref{fig:5} and \ref{fig:6} the
curves indicate that something is present around the position of
``m''. This resonance is striking by the fact that it approaches the
Fermi wavenumber when the frequency approaches zero. It might also be
appropriate here to mention that the resonances named ``a'' and ``b''
have the asymptotic value of $q_{\|}=1/n$ in the low end of the
normalized $q_{\|}$-$\omega$-spectrum, and that the resonances named
``c'' and ``d'' approaches $q_{\|}=n$ for high values of $q_{\|}/q$
and low values of $\omega/\omega_{21}$. The resonance named ``h'' is
the interband resonance.

\section{Discussion}\label{sec:V}
To give an impression of the magnitude of the phase conjugated
response, we have in Figs.~\ref{fig:3}--\ref{fig:6} highlighted the
isophotes with magnitude of $10^{20}$m$^4$/W$^2$ and
$10^{30}$m$^4$/W$^2$ by drawing them with a long-dashed curve and a
short-dashed curve, respectively. Their positions in the normalized
$q_{\|}$-$\omega$-plane shows quite clearly that most of the area
reachable within a single-mode experiment should produce a phase
conjugated response of a magnitude comparable to what one gets from
second-harmonic generation (compare Refs.~\onlinecite{Sipe:82:1,%
  Richmond:88:1,Liebsch:89:1,Heinz:91:1,Janz:93:1,Reider:95:1,%
  Pedersen:95:1,Liu:99:2,Chen:97:2}).

Knowing the positions of the resonances in the normalized
$q_{\|}$-$\omega$-plane, one could of course be tempted to plot the
magnitude of the phase conjugated response along paths following each
of the resonances (e.g., following the path of resonance ``i'', and
its continuation into ``m'') in order to give an improved
understanding of the importance of the different resonances. However,
since it would be rather difficult in an experiment to follow such a
path, and since the exact positions of the resonances probably will be
shifted in a practical situation, we have chosen not to do so. We have
in stead in Figs.~\ref{fig:8}, \ref{fig:9}, and \ref{fig:10} plotted
the intensity of the phase conjugated field along linear cuts in the
normalized $\omega$-$q_{\|}$-plane at $q_{\|}/q=0.4$, $q_{\|}/q=3.0$,
and $\omega/\omega_{21}=1.5$, respectively. Following the curves in
Figs.~\ref{fig:8}--\ref{fig:10} along their respective path on
Figs.~\ref{fig:3}--\ref{fig:6}, the appearance and dissapearance of
each resonance along the path is easily identified. From
Figs.~\ref{fig:8}--\ref{fig:10} it also appears that some of the
regions in Figs.~\ref{fig:3} and \ref{fig:4} with high density of
isophotes are zeros rather than resonances.

\begin{intextfigure}
\setlength{\unitlength}{1mm}
\psset{unit=1mm}
\begin{center}
\begin{pspicture}(-5,0)(80,95)
\put(-5,2){\epsfig{file=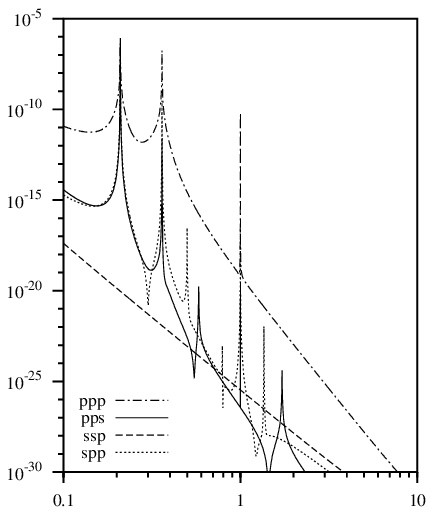,width=8.5cm}}
\rput[b]{90}(6,50){$R_{\rm{PC}}(\vec{q}_{\|},\omega)$~[m$^4$/W$^2$]}
\put(47,0){\makebox(0,0)[b]{$\omega/\omega_{21}$}}
\psframe[linecolor=white,fillstyle=solid,fillcolor=white](20,12)(38,27)
\end{pspicture}
\end{center}
\fcaption{The phase conjugation reflection coefficient is shown for the
  four combinations of polarization presented in
  Figs.~\ref{fig:3}--\ref{fig:6} in the normalized angular frequency
  range $0.1\leq\omega/\omega_{21}\leq10$ for a constant value of the
  parallel wavevector, $q_{\|}=0.4q$. Thus the four curves represents
  the ppp (dash-dot curve), pps (fully drawn curve), ssp (dashed
  curve), and spp (dotted curve) configurations.
  \label{fig:8}}
\end{intextfigure}

One of the resonances are of special interest, namely the resonance at
the interband transition frequency, which experimentally is rather
easy to tune into. Until now, resonant four-wave mixing has been
studied in other
contexts,\cite{Ducloy:84:1,Pawelek:96:1,Schirmer:97:1,Chalupszak:94:1}
but always at the point $(q_{\|},\omega)=(0,\omega_{21})$ in the
$q_{\|}$-$\omega$-plane. To go beyond that, we have plotted the phase
conjugated response in the case where the interband transition is
resonant (along the linear path in the normalized
$q_{\|}$-$\omega$-plane where $\omega=\omega_{21}$) in
Fig.~\ref{fig:11}.

\begin{intextfigure}
\setlength{\unitlength}{1mm}
\psset{unit=1mm}
\begin{center}
\begin{pspicture}(0,0)(80,90)
\put(-5,2){\epsfig{file=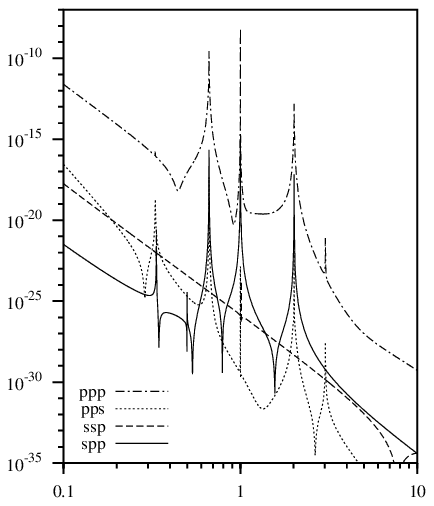,width=8.5cm}}
\rput[b]{90}(6,50){$R_{\rm{PC}}(\vec{q}_{\|},\omega)$~[m$^4$/W$^2$]}
\put(47,0){\makebox(0,0)[b]{$\omega/\omega_{21}$}}
\psframe[linecolor=white,fillstyle=solid,fillcolor=white](20,12)(38,27)
\end{pspicture}
\end{center}
\fcaption{The phase conjugation reflection coefficient is shown for the
  four combinations of polarization presented in
  Figs.~\ref{fig:3}--\ref{fig:6} in the normalized angular frequency
  range $0.1\leq\omega/\omega_{21}\leq10$ for a constant value of the
  parallel wavevector, $q_{\|}=3.0q$. The ppp configuration result is
  drawn using a dash-dot type of curve, while the pps, ssp, and spp
  configurations are drawn using dotted, dashed, and fully drawn
  curves, respectively.\label{fig:9}}
\end{intextfigure}

In configurations with only a single source field in the field-matter
interaction, such as, e.g., in linear response, second-harmonic
generation, photon drag, and photoemission the so-called self-field
approximation has proven to be quite effective. The founding argument
to use the self-field approximation is that the dynamics across the
quantum well (in the $z$-direction here) are dominating over motion in
the plane of the quantum well ($x$-$y$-plane here). Let us as a test
in the following look at the consequences of applying the self-field
approximation in the present case of degenerate four-wave mixing,
where three incident fields are present.

Working within the framework of the self-field approximation, we
observe from Eq.~\ref{eq:5} that the phase conjugated response would
have been limited to the cases where nonlinear and linear current
densities is produced in the $z$-direction. Hence, only tensor
elements with $i=z$ would contribute. Then, from Table~\ref{tab:I} we
observe that the contributions from (i) the two cases where the pump
fields have the same polarization and the probe field is s-polarized
(sss and pps), and (ii) the mixed-pump configurations spp and psp
would have been neglected. Thus, the data presented in
Figs.~\ref{fig:3} and \ref{fig:6} would have been absent. While this
is certainly a good aproximation in the pure s-polarized case, the
argument is not so good in cases with pump or probe dynamics in the
$z$-direction. Using the argument of the dominating $z$-dynamics, it
is striking that the mixed-pump configurations with s-polarized probe
field survives the self-field approximation while the two others do
not, because we would expect more dynamices in the $z$-direction from
the latter two. Another interesting conclusion is that with the loss
of Fig.~\ref{fig:3} we would also lose the resonances named ``j'',
``k'', and ``l'' in Fig.~\ref{fig:7}. At the same time we would keep
the essentially nonresonant ssp case. Comparing the raw amplitudes of
the different configurations we can see from
Figs.~\ref{fig:3}--\ref{fig:6} and \ref{fig:8}--\ref{fig:10} that in
most regions of the $q_{\|}$-$\omega$-plane, the ppp configuration
gives a response that is a few orders of magnitude larger than the
other configurations, but we also observe that the three other cases
have resonances around $q_{\|}/q=0$, while the ppp-case do not. Thus,
for near-normal incidense of the probe, the phase conjugated
reflection coefficient is larger for some of the mixed modes than for
the pure p-polarized configuration, indeed leaving room for
experiments that cannot be described within the framework of the
self-field approximation.

\begin{intextfigure}
\setlength{\unitlength}{1mm}
\psset{unit=1mm}
\begin{center}
\begin{pspicture}(-4,0)(80,91.5)
\put(-5,2){\epsfig{file=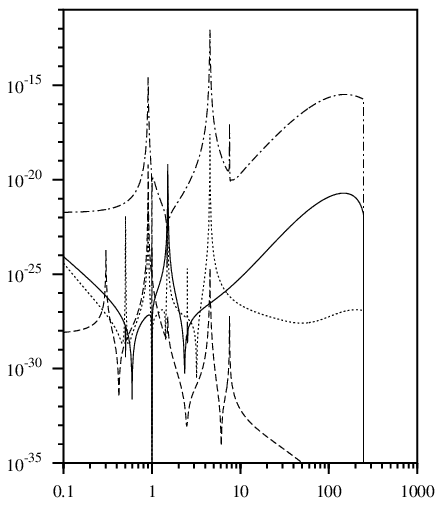,width=8.5cm}}
\put(-0,2){\epsfig{file=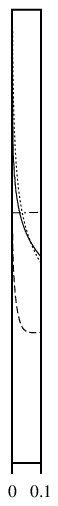,width=0.828cm}}
\rput[b]{90}(0,50){$R_{\rm{PC}}(\vec{q}_{\|},\omega)$~[m$^4$/W$^2$]}
\put(47,0){\makebox(0,0)[b]{$q_{\|}/q$}}
\psline[linewidth=0.25mm]{>-<}(69.1,4)(69.1,91.5)
\put(69,0){\makebox(0,0)[b]{$k_F/q$}}
\end{pspicture}
\end{center}
\fcaption{The phase conjugation reflection coefficient is drawn on a
  logarithmic scale for the four combinations of polarization
  presented in Figs.~\ref{fig:3}--\ref{fig:6} in the normalized
  parallel wavevector range $0\leq{}q_{\|}/q\leq{}k_{F}/q$ for a
  constant value of the angular frequency, $\omega=1.5\omega_{21}$. In
  the strip to the left, the abscissa is linear, while it is
  logarithmic in the right part of the figure. The scale of the
  ordinate is the same in both frames. The upper curve (dash-dot)
  shows the result for the ppp configuration of polarizations, while
  the dashed curve shows the pps result, the fully drawn curve shows
  the ssp result, and the dotted curve shows the spp
  result.\label{fig:10}}
\end{intextfigure}

All in all, we may conclude from the above discussion that although
the argument behind the self-field approximation remains intact, when
one allows more than one incident field to participate in the
interaction (as in, e.g., sum- and difference frequency generation, or
degenerate four-wave mixing), one should be careful in applying the
self-field approximation in cases where mixed polarizations of the
incident fields are allowed.

Outside the resonances the influence of the relaxation time is
insignificant, but around the resonances the choice of relaxation time
has a great influence on the width (in the $q_{\|}$-space) and
amplitude of each resonance. Choosing adequate relaxation times
$\tau_{nm}$ is a difficult problem and it appears from
Fig.~\ref{fig:12} how big impact the relaxation time has on the phase
conjugation reflection coefficient. The intraband relaxation time in
the occupied state ($\tau_{11}$) has been chosen in accordance with
Ref.~\onlinecite{Andersen:98:2} to be 3fs. For the unoccupied state
the relaxation time $\tau_{22}$ (see Fig.~\ref{fig:2}) has been chosen
to approach infinity. In the present case where also interband
transitions contribute to the phase conjugated response, the intraband
relaxation time is of little importance, and thus it is the choice of
interband relaxation times (here $\tau_{21}$ and $\tau_{12}$) that are
critical.  In the present calculation we assume no relaxation from
state $|1\rangle$ to state $|2\rangle$, letting
$\tau_{12}\rightarrow\infty$.

\begin{intextfigure}
\setlength{\unitlength}{1mm}
\psset{unit=1mm}
\begin{center}
\begin{pspicture}(-4,0)(80,91.5)
\put(-5,2){\epsfig{file=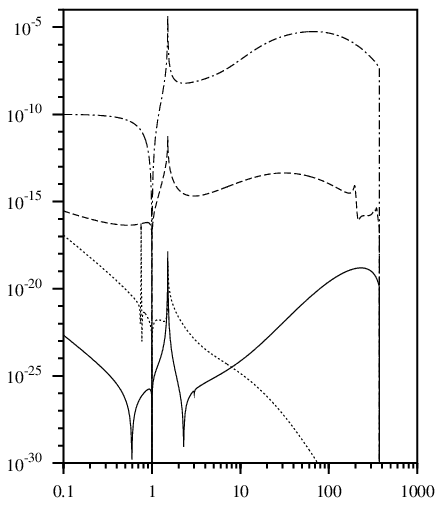,width=8.5cm}}
\put(-0,2){\epsfig{file=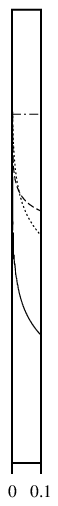,width=0.828cm}}
\rput[b]{90}(0,50){$R_{\rm{PC}}(\vec{q}_{\|},\omega)$~[m$^4$/W$^2$]}
\put(47,0){\makebox(0,0)[b]{$q_{\|}/q$}}
\psline[linewidth=0.25mm]{>-<}(71.75,4)(71.75,91.5)
\put(71.75,0){\makebox(0,0)[b]{$k_F/q$}}
\end{pspicture}
\end{center}
\fcaption{The phase conjugation reflection coefficient is shown on a
  logarithmic scale for the four combinations of polarization
  presented in Figs.~\ref{fig:3}--\ref{fig:6} in the normalized
  parallel wavevector range $0\leq{}q_{\|}/q\leq{}k_{F}/q$ when the
  value of the angular frequency is exactly equal to the interband
  transition frequency, $\omega=\omega_{21}$. As in Fig.~\ref{fig:10},
  the strip to the left shows the range $0\leq{}q_{\|}/q\leq0.1$ with
  linear abscissa, while the rest is plotted with logarithmic
  abscissa. The scale of the ordinate is the same for both frames. In
  this figure, the dash-dot curve corresponds to the ppp case as in
  the previous figure, but the dotted curve to the pps result. The
  fully drawn curve corresponds to the ssp case as before, and the
  dashed curve to the spp result.\label{fig:11}}
\end{intextfigure}

The phase conjugation reflection coefficent has in Fig.~\ref{fig:12}
been plotted for four values of the relaxation time from state
$|2\rangle$ to state $|1\rangle$, namely (i) 30fs and (ii) 200fs,
which are typical values one would find for bulk
copper\cite{Ashcroft:76:1} at (i) room temperature and (ii) at 77K,
(iii) 3fs, and (iv) 2ps. The value in case (iii) is obtained by a
conjecture based on the difference between measured data for a lead
quantum well\cite{Jalochowski:97:1} and the bulk value for lead at
room temperature. The difference between the relaxation times measured
by Jalochowski, Str{\.o}{\.z}ak, and Zdyb\cite{Jalochowski:97:1} is
for two monolayers approximately one order of magnitude. Case (iv) is
included to see the effect of raising the value of the relaxation time
one order of magnitude, thus essentially assuming a better conductance
than in case (ii). The values (i)--(iii) are the same values as we
chose in our description of the single-level quantum-well case where
only intraband transitions were allowed,\cite{Andersen:98:2} but since
the interband transition is of a more bulk-like character we have in
the present calculations chosen $\tau_{21}=200$fs. We notice that in
the case where both pump fields are s-polarized (polarized in the
plane of the quantum well), the phase conjugated response does not
vary as a function of the interband relaxation time, whereas in the
other three cases the general tendency is that they have larger
magnitudes for larger values of the relaxation time.

\begin{intextfigure}
\setlength{\unitlength}{1mm}
\psset{unit=1mm}
\begin{center}
\begin{pspicture}(-0,0)(80,110)
\put(-5,80){\epsfig{file=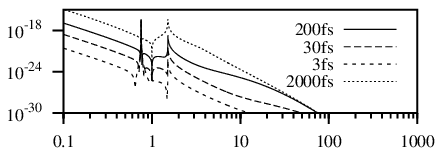,width=8.5cm}}
\put(-5,54){\epsfig{file=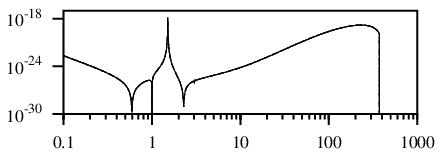,width=8.5cm}}
\put(-5,28){\epsfig{file=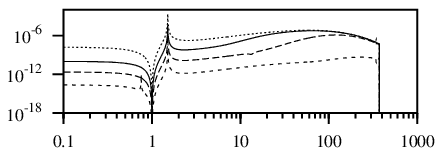,width=8.5cm}}
\put(-5,2){\epsfig{file=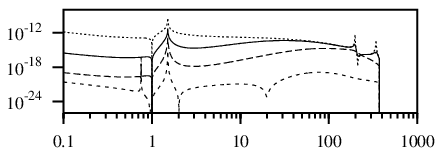,width=8.5cm}}
\rput[b]{90}(6,54){$R_{\rm{PC}}(\vec{q}_{\|},\omega)$~[m$^4$/W$^2$]}
\put(47,0){\makebox(0,0)[b]{$q_{\|}/q$}}
\put(19,90){\makebox(0,0)[bl]{pps}}
\put(19,64){\makebox(0,0)[bl]{ssp}}
\put(19,38){\makebox(0,0)[bl]{ppp}}
\put(19,12){\makebox(0,0)[bl]{spp}}
\psframe[linecolor=white,fillstyle=solid,fillcolor=white](55.7,93.2)(77,106)
\psline[linewidth=0.25mm]{>-<}(71.75,4)(71.75,109.5)
\put(71.75,0){\makebox(0,0)[b]{$k_F/q$}}
\end{pspicture}
\end{center}
\fcaption{The phase conjugation reflection coefficient is shown for
  interband transition resonance for different values of the interband
  relaxation time $\tau_{21}\in\{3,30,200\}$ femtoseconds, and $2$
  picoseconds.  The fully drawn curve corresponds to $200$fs, the
  long-dashed curve to $30$fs, the short-dashed curve to $3$fs, and
  the dotted curve to $2$ps.\label{fig:12}}
\end{intextfigure}

\section{Conclusions}\label{sec:VI}
Our main conclusion from this work is that DFWM in a thin metallic
film gives rise to several resonance structures even in the
propagating regime of the $q_{\|}$-spectrum. Furthermore the coupling
by the phase conjugation reflection coefficient is of a magnitude that
is well within experimental reach. Thus, also single mode excitation
in the experimentally feasible regime (up to around $n=3$) should be
possible by use of the standard Otto\cite{Otto:68:1} or
Kretschmann\cite{Kretschmann:68:1} techniques, and a qualitative
comparison with the present work should be possible.  However, for a
better quantitative comparison in a specific system, it will be
necessary to refine the numerical calculation by, e.g., abandoning the
IB model in favor of one of the flavors of the KKR, LAPW or LMTO
models, although such a task may prove to be strenuous.
\endmulticols

\appendix
\section{Nonlinear conductivity tensor}\label{app:DFWMtensor}\label{app:A}
Under the assumption that the electron dynamics is free-electron-like
in the plane of the quantum well the nonlinear response function
$\tensor{\Xi}(z,z';\vec{q}_{\|},\vec{k}_{\|},\omega)$ [given by
Eq.~(\ref{eq:36}), and with tensor elements $\Xi_{ijkh}$] can be
obtained from the results established for
$\tensor{\Xi}^{\rm{G}}(z,z',z'',z''';\vec{q}_{\|},\vec{k}_{\|},\omega)$ in
Ref.~\onlinecite{Andersen:97:1}. Upon integration over $z''$ and
$z'''$ [and use of Eq.~(\ref{eq:37})] one gets
\FL
\begin{eqnarray}
\lefteqn{
 {\Xi}_{ijkh}(z,z';\vec{q}_{\|},\vec{k}_{\|},\omega)=
 -{1\over8\hbar^3}{1\over(2\pi)^2}
 {2\over(i\omega)^3}\sum_{nmvl}\int
 {1\over\tilde{\omega}_{nm}(\vec{\kappa}_{\|}+\vec{q}_{\|},\vec{\kappa}_{\|})
 -\omega}
}\nonumber\\ &\quad&\times
\left\{
 \left[
 \left({f_{l}(\vec{\kappa}_{\|}-\vec{k}_{\|})
 -f_{m}(\vec{\kappa}_{\|})\over\tilde{\omega}_{l{}m}(\vec{\kappa}_{\|}
 -\vec{k}_{\|},\vec{\kappa}_{\|})-\omega}
 +{f_{l}(\vec{\kappa}_{\|}-\vec{k}_{\|})-f_{v}(\vec{\kappa}_{\|})
 \over\tilde{\omega}_{vl}(\vec{\kappa}_{\|},
 \vec{\kappa}_{\|}-\vec{k}_{\|})-\omega}\right)
 {1\over\tilde{\omega}_{v{}m}(\vec{\kappa}_{\|},\vec{\kappa}_{\|})-2\omega}
 +\left({f_{l}(\vec{\kappa}_{\|}-\vec{k}_{\|})
 -f_{v}(\vec{\kappa}_{\|})\over
 \tilde{\omega}_{vl}(\vec{\kappa}_{\|},
 \vec{\kappa}_{\|}-\vec{k}_{\|})-\omega}
\right.\right.\right.\nonumber\\ &&\left.\left.
 +{f_{n}(\vec{\kappa}_{\|}+\vec{q}_{\|})-f_{v}(\vec{\kappa}_{\|})\over
 \tilde{\omega}_{nv}(\vec{\kappa}_{\|}+\vec{q}_{\|},\vec{\kappa}_{\|})
 +\omega}\right)
 {1\over\tilde{\omega}_{nl}(\vec{\kappa}_{\|}+\vec{q}_{\|},
 \vec{\kappa}_{\|}-\vec{k}_{\|})}
\right]
\nonumber\\ &&\times
 \int{}j_{h,ml}(z''';2\vec{\kappa}_{\|}-\vec{k}_{\|})dz'''
 \int{}j_{k,lv}(z'';2\vec{\kappa}_{\|}-\vec{k}_{\|})dz''
 j_{j,v{}n}(z';2\vec{\kappa}_{\|}+\vec{q}_{\|})
\nonumber\\ && 
 +
 \left[
 \left({f_{l}(\vec{\kappa}_{\|}-\vec{k}_{\|})
 -f_{m}(\vec{\kappa}_{\|})\over\tilde{\omega}_{l{}m}(\vec{\kappa}_{\|}
 -\vec{k}_{\|},\vec{\kappa}_{\|})-\omega}
 +{f_{l}(\vec{\kappa}_{\|}-\vec{k}_{\|})-f_{v}(\vec{\kappa}_{\|}
 -\vec{k}_{\|}+\vec{q}_{\|})\over
 \tilde{\omega}_{vl}(\vec{\kappa}_{\|}-\vec{k}_{\|}+\vec{q}_{\|},
 \vec{\kappa}_{\|}-\vec{k}_{\|})+\omega}\right)
 {1\over\tilde{\omega}_{v{}m}(\vec{\kappa}_{\|}-\vec{k}_{\|}
 +\vec{q}_{\|},\vec{\kappa}_{\|})}
\right.\nonumber\\ &&\left.
 +\left({f_{l}(\vec{\kappa}_{\|}-\vec{k}_{\|})
 -f_{v}(\vec{\kappa}_{\|}-\vec{k}_{\|}+\vec{q}_{\|})\over
 \tilde{\omega}_{vl}(\vec{\kappa}_{\|}-\vec{k}_{\|}+\vec{q}_{\|},
 \vec{\kappa}_{\|}-\vec{k}_{\|})+\omega}
 +{f_{n}(\vec{\kappa}_{\|}+\vec{q}_{\|})
 -f_{v}(\vec{\kappa}_{\|}-\vec{k}_{\|}+\vec{q}_{\|})\over
 \tilde{\omega}_{nv}(\vec{\kappa}_{\|}+\vec{q}_{\|},\vec{\kappa}_{\|}
 -\vec{k}_{\|}+\vec{q}_{\|})-\omega}
 \right)
 {1\over\tilde{\omega}_{nl}(\vec{\kappa}_{\|}
 +\vec{q}_{\|},\vec{\kappa}_{\|}-\vec{k}_{\|})}
 \right]
\nonumber\\ &&\times
 \int{}j_{h,ml}(z''';2\vec{\kappa}_{\|}-\vec{k}_{\|})dz'''
 \int{}j_{k,v{}n}(z'';2\vec{\kappa}_{\|}-\vec{k}_{\|}+2\vec{q}_{\|})dz''
 j_{j,lv}(z';2\vec{\kappa}_{\|}-2\vec{k}_{\|}+\vec{q}_{\|})
\nonumber\\ &&
 +
 \left[
 \left({f_{l}(\vec{\kappa}_{\|}+\vec{q}_{\|})-f_{m}(\vec{\kappa}_{\|})
 \over\tilde{\omega}_{l{}m}(\vec{\kappa}_{\|}+\vec{q}_{\|},
 \vec{\kappa}_{\|})+\omega}
 +{f_{l}(\vec{\kappa}_{\|}+\vec{q}_{\|})-f_{v}(\vec{\kappa}_{\|}
 -\vec{k}_{\|}+\vec{q}_{\|})\over
 \tilde{\omega}_{vl}(\vec{\kappa}_{\|}-\vec{k}_{\|}+\vec{q}_{\|},
 \vec{\kappa}_{\|}+\vec{q}_{\|})-\omega}\right)
 {1\over\tilde{\omega}_{v{}m}(\vec{\kappa}_{\|}-\vec{k}_{\|}
 +\vec{q}_{\|},\vec{\kappa}_{\|})}
\right.\nonumber\\ &&\left.
 +\left({f_{l}(\vec{\kappa}_{\|}+\vec{q}_{\|})
 -f_{v}(\vec{\kappa}_{\|}-\vec{k}_{\|}+\vec{q}_{\|})\over
 \tilde{\omega}_{vl}(\vec{\kappa}_{\|}-\vec{k}_{\|}+\vec{q}_{\|},
 \vec{\kappa}_{\|}+\vec{q}_{\|})-\omega}
 +{f_{n}(\vec{\kappa}_{\|}+\vec{q}_{\|})
 -f_{v}(\vec{\kappa}_{\|}-\vec{k}_{\|}+\vec{q}_{\|})\over
 \tilde{\omega}_{nv}(\vec{\kappa}_{\|}+\vec{q}_{\|},\vec{\kappa}_{\|}
 -\vec{k}_{\|}+\vec{q}_{\|})-\omega}\right)
 {1\over\tilde{\omega}_{nl}(\vec{\kappa}_{\|}
 +\vec{q}_{\|},\vec{\kappa}_{\|}+\vec{q}_{\|})-2\omega}
 \right]
\nonumber\\ &&\times\left.
 \int{}j_{h,lv}(z''';2\vec{\kappa}_{\|}-\vec{k}_{\|}+2\vec{q}_{\|})dz'''
 \int{}j_{k,v{}n}(z'';2\vec{\kappa}_{\|}-\vec{k}_{\|}+2\vec{q}_{\|})dz''
 j_{j,ml}(z';2\vec{\kappa}_{\|}+\vec{q}_{\|})
\right\}
 j_{i,nm}(z;2\vec{\kappa}_{\|}+\vec{q}_{\|})
 d^2\kappa_{\|}.
\label{eq:A5}
\end{eqnarray}
\narrowtext\noindent 

\section{On the solution to the integrals over $\vec{\kappa}_{\|}$ in
  the low-temperature limit}\label{app:B}
In this appendix we discuss how analytical solutions to the integrals
over the electronic wavevector, $\vec{\kappa}_{\|}$, appearing in the
linear and nonlinear conductivity tensor may be obtained, and for
simplicity the discussion is limited to cover the low-temperature
limit. These integrals can, when scattering takes place in the
$x$-$z$-plane, be expressed as a sum over terms of the general type
\begin{equation}
 {\cal{F}}_{pq}^{\beta}(n,\{a\},\{b\},s)=
 \int_{-\infty}^{\infty}\int_{-\infty}^{\infty}
 {{\kappa}_{x}^{p}\kappa_{y}^{q}f_{n}(\vec{\kappa}_{\|}+s\vec{e}_{x})
 \over\prod_{k=1}^{\beta}[a_k\kappa_{x}+b_k]}
 d\kappa_{x}d\kappa_{y},
\label{eq:B1}
\end{equation}
where $p,k,\beta$ are nonnegative integers, and $q$ is an even
nonnegative integer. The functions depends on (i) the quantum number
$n$, which is a positive nonzero integer, (ii) a set of real
quantities, $\{a\}\equiv\{a_1,\dots,a_\beta\}$ appearing in front of
the integration variable $\kappa_x$ in the denominator, (iii) a set of
complex nonzero quantities, $\{b\}\equiv\{b_1,\dots,b_\beta\}$
appearing also in the denominator, and (iv) the real quantity s
representing the displacement (in the $x$-direction) of the center of
the Fermi-Dirac distribution function from
$(\kappa_x,\kappa_y)=(0,0)$.  The quantity s together with each
element in the set $\{a\}$ are in general functions of the parallel
components of the probe and pump wavevectors, $\vec{q}_{\|}$ and
$\vec{k}_{\|}$. Each element in the set $\{b\}$ is furthermore a
function of $\tau$, the relaxation time.

The combinations of $p$ and $q$ needed in Eq.~(\ref{eq:B1}) in order
to solve the integrals over $\vec{\kappa}_{\|}$ in the nonlinear
conductivity tensor are $(p,q)\in\{(0,0)$, $(0,2)$, $(0,4)$, $(1,0)$,
$(1,2)$, $(2,0)$, $(2,2)$, $(3,0)$, $(4,0)\}$, and
$\beta\in\{1,2,3\}$. However, functions with $\beta=2$ and $\beta=3$
can be expressed in terms of functions with $\beta=1$ in the following
way:
\widetext
\begin{eqnarray}
 {\cal{F}}_{pq}^{2}(n,a_1,a_2,b_1,b_2,s)&=&
 {a_1{\cal{F}}_{pq}^{1}(n,a_1,b_1,s)-a_2{\cal{F}}_{pq}^{1}(n,a_2,b_2,s)
  \over{}a_1b_2-a_2b_1}, 
\label{eq:B3}
\\
 {\cal{F}}_{pq}^{3}(n,a_1,a_2,a_3,b_1,b_2,b_3,s)&=&
 {a_1^2{\cal{F}}_{pq}^{1}(n,a_1,b_1,s)
  \over(a_2b_1-b_2a_1)(a_3b_1-b_3a_1)}
 \!+{a_2^2{\cal{F}}_{pq}^{1}(n,a_2,b_2,s)
  \over(a_2b_1-b_2a_1)(a_3b_2-b_3a_2)}
 +{a_3^2{\cal{F}}_{pq}^{1}(n,a_3,b_3,s)
  \over(a_3b_1-b_3a_1)(a_3b_2-b_3a_2)}.
\nonumber\\
\label{eq:B4}
\end{eqnarray}
\multicols{2}
\noindent
Eqs.~(\ref{eq:B3}) and (\ref{eq:B4}) are given with the provision that
the values of the different $a_k$ are nonzero, $k\in\{1,2\}$ in
Eq.~(\ref{eq:B3}) and $k\in\{1,2,3\}$ in Eq.~(\ref{eq:B4}). If any
$a_k$, for instance $a_1$, becomes zero, we see from Eq.~(\ref{eq:B1})
that the order (in $\kappa_x$) of the denominator becomes smaller by
one. This implies that
${\cal{F}}_{pq}^{2}(n,0,a_2,b_1,b_2,s)={\cal{F}}_{pq}^{1}(n,a_2,b_2,s)/b_1$
in Eq.~(\ref{eq:B3}).  The similar conclusion for Eq.~(\ref{eq:B4}) is
${\cal{F}}_{pq}^{3}(n,0,a_2,a_3,b_1,b_2,b_3,s)={\cal{F}}_{pq}^{2}(n,a_2,a_3,b_2,b_3,s)/b_1$.
Analogous reductions applies for any other $a_k=0$.

In the low-temperature limit the Fermi-Dirac distribution function is
zero outside the Fermi sphere and equal to one inside, and it is
therefore advantageous to shift $\kappa_x$ by $-s$, and afterwards
carry out the integrations in polar $(r,\theta)$ coordinates. Using
$\kappa_x=r\cos\theta$, $\kappa_y=r\sin\theta$, and 
$d\kappa_xd\kappa_y=rd\theta{}dr$, the integrals to be solved are of
the type
\begin{equation}
 {\cal{F}}_{pq}^{1}(n,a,b,s)
 =\int_{0}^{{\alpha}(n)}\int_{0}^{2\pi}{r(r\cos\theta-s)^{p}(r\sin\theta)^{q}
  \over{}b-as+ar\cos\theta}d\theta dr,
\label{eq:B5}
\end{equation}
dropping the now superfluous index on $a$ and $b$. The upper limit of
the radial integration is $\alpha(n)=\sqrt{k_F^2-(\pi{}n/d)^2}$,
$k_F>\pi{}n/d$. If $k_F<\pi{}n/d$, the Fermi-Dirac distribution
function is zero, and thus the integral vanishes. Physically,
$\alpha(n)$ may be characterized as the two-dimensional Fermi
wavenumber for electrons in subband $n$.

Since the following treatment is a formal solution of
Eq.~(\ref{eq:B5}), we will also drop the reference to $n$ for brevity,
letting $\alpha\equiv\alpha(n)$. To solve Eq.~(\ref{eq:B5}), let us
make the substitutions
\begin{equation}
 \eta\equiv{b-as\over{}a\alpha},\quad 
 r\equiv\alpha{}u,
\label{eq:B6}
\end{equation}
and thereby turn Eq.~(\ref{eq:B5}) into
\begin{eqnarray}
\lefteqn{
 {\cal{F}}_{pq}^{1}(\alpha,\eta,s)=
}\nonumber\\ &\quad&
 {\alpha^q\over{}a}\int_0^1\int_0^{2\pi}
 {u^{q+1}(\alpha{}u\cos\theta-s)^p(1-\cos^2\theta)^{q/2}
 \over\eta+u\cos\theta}d\theta{}du,
\nonumber\\
\label{eq:B7}
\end{eqnarray}
i.e., compared to the possible values of $p$ and $q$, an expression
where the angular integral is expressed as a sum of terms of the form
$\cos^h\theta$ in the nominator, where $h\in\{0,1,2,3,4\}$. To carry
out the angular integrals we put $t=\exp(i\theta)$ so that the
integrals become of the type
\begin{equation}
 \int_0^{2\pi}{\cos^h\theta\over\eta+u\cos\theta}d\theta
 ={1\over{}2^hiu}\oint{(1+t^2)^h\over{}t^h(t-t_+)(t-t_-)}dt.\!\!
\label{eq:B17}
\end{equation}
In Eq.~(\ref{eq:B17}), the poles at $t_{\pm}$ in the $t$-plane are
located at
\begin{equation}
 t_{\pm}=-{\eta\over{}u}\pm\sqrt{\left({\eta\over{}u}\right)^2-1},
\label{eq:B18}
\end{equation}
and the integration runs along the unit circle. Since $t_+t_-=1$, one
of these poles is located inside the unit circle while the other is
outside. When $h>0$ there is an additional pole of order $h$ at $t=0$.
Using the unit circle as contour, residue calculations give the
nontrivial solutions
\begin{eqnarray}
 \int_{0}^{2\pi}{1\over\eta+u\cos\theta}d\theta
 &=&{2\pi\over\sqrt{\eta^2-u^2}},
\label{eq:B19}\\
 \int_{0}^{2\pi}{\cos\theta\over\eta+u\cos\theta}d\theta
 &=&{2\pi\over{}u}\left[1-{\eta\over\sqrt{\eta^2-u^2}}\right],
\label{eq:B20}\\
 \int_{0}^{2\pi}{\cos^2\theta\over\eta+u\cos\theta}d\theta
 &=&{2\pi\eta\over{}u^2}\left[{\eta\over\sqrt{\eta^2-u^2}}-1\right],
\label{eq:B21}\\
 \int_{0}^{2\pi}{\cos^3\theta\over\eta+u\cos\theta}d\theta
 &=&{\pi\over{}u}+{2\pi\eta^2\over{}u^3}
 \left[1-{\eta\over\sqrt{\eta^2-u^2}}\right],
\label{eq:B22}\\
 \int_{0}^{2\pi}{\cos^4\theta\over\eta+u\cos\theta}d\theta
 &=&{2\pi\eta^3\over{}u^4}\left[{\eta\over\sqrt{\eta^2-u^2}}-1\right]
 -{\pi\eta\over{}u^2}.
\label{eq:B23}
\end{eqnarray}
To finish the formal solution, (i) insert these results into
Eq.~(\ref{eq:B7}), (ii) carry out the elementary radial integrations
(see, e.g., Ref.~\onlinecite{Gradshteyn:94:1}, Sec.~2.27), (iii)
backsubstitute $\eta$, and (iv) check convergence for $a\rightarrow0$.
Step (iv) can be carried out by use of a binomial series expansion of
the square roots appearing, and a comparison the the result one gets
by setting $a=0$ already in Eq.~(\ref{eq:B5}). The solution to the
integrals appearing in Eqs.~(\ref{eq:7}) and (\ref{eq:A5}) are then
found in a straightforward manner, but since the algebraic expressions
are rather long, we will omit presenting them here [they can be found
in Ref.~\onlinecite{Andersen:98:1} together with explicit expressions
for the case where $a=0$].

\section{Denominator coefficients}\label{app:C}
\begin{eqnarray}
 a_1&=&{\hbar{}k_{\|}\over{}m_e},
\label{eq:C1}\\
 a_2&=&{\hbar{}q_{\|}\over{}m_e},
\label{eq:C2}\\
 a_3&=&{\hbar\over{}m_e}(q_{\|}+k_{\|}),
\label{eq:C3}\\
 a_4&=&{\hbar\over{}m_e}(q_{\|}-k_{\|}),
\label{eq:C4}\\
 b_{nm}^1&=&{1\over\hbar}(\varepsilon_n-\varepsilon_m)
 +{\hbar{}k_{\|}^2\over2m_e}-\omega-{\rm{i}}\tau_{nm}^{-1},
\label{eq:C5}\\
 b_{nm}^2&=&{1\over\hbar}(\varepsilon_n-\varepsilon_m)
 -{\hbar{}k_{\|}^2\over2m_e}-\omega-{\rm{i}}\tau_{nm}^{-1},
\label{eq:C6}\\
 b_{nm}^3&=&{1\over\hbar}(\varepsilon_n-\varepsilon_m)
 +{\hbar{}q_{\|}^2\over2m_e}+\omega-{\rm{i}}\tau_{nm}^{-1},
\label{eq:C7}\\
 b_{nm}^4&=&{1\over\hbar}(\varepsilon_n-\varepsilon_m)
 +{\hbar{}q_{\|}^2\over2m_e}-\omega-{\rm{i}}\tau_{nm}^{-1},
\label{eq:C8}\\
 b_{nm}^5&=&{1\over\hbar}(\varepsilon_n-\varepsilon_m)
 +{\hbar\over2m_e}(q_{\|}-k_{\|})^2-{\rm{i}}\tau_{nm}^{-1},
\label{eq:C9}\\
 b_{nm}^6&=&{1\over\hbar}(\varepsilon_n-\varepsilon_m)
 +{\hbar\over2m_e}(q_{\|}^2-k_{\|}^2)-{\rm{i}}\tau_{nm}^{-1},
\label{eq:C10}\\
 b_{nm}^7&=&{1\over\hbar}(\varepsilon_n-\varepsilon_m)
 +{\hbar{}q_{\|}\over2m_e}(q_{\|}-2k_{\|})+\omega-{\rm{i}}\tau_{nm}^{-1},
\label{eq:C11}\\ 
 b_{nm}^8&=&{1\over\hbar}(\varepsilon_n-\varepsilon_m)
 +{\hbar{}k_{\|}\over2m_e}(2q_{\|}-k_{\|})-\omega-{\rm{i}}\tau_{nm}^{-1},
\label{eq:C12}\\
 b_{nm}^9&=&{1\over\hbar}(\varepsilon_n-\varepsilon_m)
 +{\hbar{}k_{\|}\over2m_e}(k_{\|}-2q_{\|})-\omega-{\rm{i}}\tau_{nm}^{-1}.
\label{eq:C13}
\end{eqnarray}
\end{multicols}

\vskip1cm plus 0.5mm minus 0.5mm
\hbox to\hsize{\hfil\rule{7cm}{.1mm}\hfil}%
\vskip0.3cm plus 0.5mm minus 0.5mm
\begin{multicols}{2}

\end{multicols}

\end{document}